\documentclass[preprint,12pt]{elsarticle}

\usepackage{amssymb}
\usepackage{amsmath,amsfonts,amssymb}
\usepackage{graphicx,color}

\newcommand{\X}{\mathbf X}
\newcommand{\x}{\mathbf x}
\renewcommand{\u}{\mathbf u}
\renewcommand{\v}{\mathbf v}
\newcommand{\w}{\mathbf w}
\newcommand{\f}{\mathbf f}
\newcommand{\F}{\mathbf F}
\newcommand{\q}{\mathbf q}
\newcommand{\J}{\mathbf J}
\newcommand{\T}{\mathbf T}
\newcommand{\D}{\mathbf D}

\newcommand{\real}{\mathbb R}
\newcommand{\bs}{\boldsymbol}
\newcommand{\mb}{\mathbf}
\newcommand{\dps}{\displaystyle}

\journal{Computational Materials Science}

\begin{document}

\begin{frontmatter}



\title{Calculation of Cauchy stress tensor in molecular dynamics system
with a generalized Irving-Kirkwood formulism}


\author[whu]{Jerry Zhijian Yang}
\author[whu]{Shukai Du}

\address[whu]{School of Mathematics and Statistics, Wuhan University, P.R. of China.}

\begin{abstract}
Irving and Kirkwood \cite{irving1950statistical} formulism (IK formulism) provides a way to compute continuum mechanics quantities at certain location in terms of molecular variables. To make the approach more practical in computer simulation, Hardy \cite{hardy1982formulas} proposed to use a spacial kernel function that couples continuum quantities with atomistic information. To reduce irrational fluctuations, Murdoch \cite{murdoch1994continuum} proposed to use a temporal kernel function to smooth the physical quantities obtained in Hardy's approach. In this paper, we generalize the original IK formulism to systematically incorporate both spacial and temporal average. The Cauchy stress tensor is derived in this generalized IK formulism (g-IK formulism). Analysis is given to illuminate the connection and difference between g-IK formulism and traditional temporal post-process approach. The relationship between Cauchy stress and first Piola-Kirchhoff stress is restudied in the framework of g-IK formulism. Numerical experiments using molecular dynamics are conducted to examine the analysis results.
\end{abstract}


\end{frontmatter}







\section{Introduction}
One commonly existing problem in material science is the time-scale
and/or length-scale gap between continuum mechanics and molecular dynamics methods of studying the system
\cite{LiYaE09,li2005multiscale,LuHyMoBeRo06,NiRoSh06,NiChERo04,ReE05}. To overcome
this difficulty, one significant work is to develop suitable definitions for continuum variables that are calculable
within an atomistic system.

Lots of work has been done in this direction.
The earliest work of defining stress in expression of microscopic quantities could date back to Cauchy in the
1820 with his aim to define stress in a crystalline solid \cite{cauchy1828equilibre,cauchy1828pression}.
Cauchy's original definition emerges from the intuitive idea of identifying stress with the force per unit area carried by the bonds that cross a given surface.
Tsai \cite{tsai1979virial} in 1979 extended the definition given by Cauchy to finite
temperature by taking into consideration the momentum flux passing through the surface.
However, since their approaches are tied to some particular surface being considered, they actually constitute definitions for traction but not for stress tensor.

The first definition of stress as a
tensorial quantity was in the late 19th century, when Clausius \cite{Clausius1870} and Maxwell \cite{Maxwell1864xlv} developed the virial theorem to calculate stress in a homogeneous system. After that, Irving and Kirkwood
\cite{irving1950statistical} derived
expressions for local stress tensor and heat current density in terms of molecular degrees of freedom, which extends virial theorem to non-homogeneous system. However, due to the Dirac Delta function used by Irving and Kirkwood in definitions of continuum quantities, numerical implementation is not straightforward. Hardy \cite{hardy1982formulas}
is one of the first to employ
finite-valued and finite-ranged localization function, called kernel function, to replace the Dirac Delta function
initially used in IK formulism. Hardy's work makes it possible to construct a self-consistent and practical manner
of distributing discrete atomic contributions to thermal mechanical fields.

Hardy's work has well coupled continuum quantities with atomistic information in spacial aspect. Otherwise, due to the
inconsistency of time-scales between the two systems, the thermal mechanical fields obtained this way still exhibit irrational fluctuations.
Murdoch and Bedeaux \cite{murdoch1993physical,murdoch1994continuum} studied how temporal average can be done after spacial average is obtained.
However, as the temporal average is not part of the original IK formulism, a careless usage of it may lead to violation of conservation laws (as we will show, the time averaged Hardy stress violates the conservation of momentum).

In this paper, we generalize the original IK formulism's definition of Dirac Delta function to a space-time coupled kernel function (space-time kernel). Through this method, we can overcome the fluctuations caused by inconsistency of space/time-scales, and also keep the work in a consistent manner of the original IK formulism.
The following issues will be in our main concerns:
\begin{itemize}
\item
Unlike traditional way of making temporal post-process on physical quantities, g-IK formulism aims at incorporating spacial and temporal average in a uniform way with a space-time kernel function. This approach is a generalization of temporal post-process (separable kernel case) method, for it gives more choices (inseparable kernel case) of calculating continuum fields from atomistic quantities.
\item
When the kernel function used in g-IK formulism is space-time separable (see section \ref{sec:g-IK-t-tpp} for the definition), we show the Cauchy stress tensor derived in g-IK formulism (g-IK stress) is equivalent to the stress derived in Murdoch' paper \cite{murdoch1994continuum}. This stress is different from the time-averaged Hardy stress. We give an analysis on the difference between g-IK stress and time-averaged Hardy stress and show how the difference rely on the spacial (or temporal) radius of the space-time kernel function. Some numerical experiments are carried to estimate the difference under different spacial (or temporal) radius.
\item
The classical relationship (${\bs\sigma}={\det(\mb F)}^{-1}\mb F\cdot{\mb P}$) between Cauchy stress and first Piola-Kirchhoff (PK) stress is restudied in the framework of g-IK formulism. When the kernel function is in uniform-weighting form, an inequality is given to estimate $\big\|{\bs\sigma}-{\det(\mb F)}^{-1}\mb F\cdot{\mb P}\big\|_2$, which turns out to rely on system temperature ($T$) and temporal radius of the kernel function ($r_t$). When the kernel function is in general form, we give an analysis on the classical relationship (${\bs\sigma}={\det(\mb F)}^{-1}\mb F\cdot{\mb P}$) when the system in zero temperature. Numerical experiments are also carried to estimate $\big\|{\bs\sigma}-{\det(\mb F)}^{-1}\mb F\cdot{\mb P}\big\|_2$ under different temporal radius and system temperatures.
\end{itemize}

The rest of the paper is organized as follows: in section \ref{sec:atom-to-cont}, we review the definitions of virial stress and the derivation of Hardy stress. In section \ref{sec:generalized-IK}, we discuss the rational definitions for mass density and momentum and the derivation of Cauchy stress tensor in g-IK formulism. In section \ref{sec:g-IK-t-tpp}, we first show that the g-IK stress is equal to Murdoch's stress \cite{murdoch1994continuum} when the kernel used in g-IK formulism is space-time separable, then we show that there lies an intrinsic difference between g-IK stress and time-averaged Hardy stress and give an analysis on the difference. In section \ref{sec:pk-cauchy-in-gIK}, we study the classical relationship (${\bs\sigma}={\det(\mb F)}^{-1}\mb F\cdot{\mb P}$) between PK and Cauchy stress in the framework of g-IK formulism, an inequality is given to estimate $\big\|{\bs\sigma}-{\det(\mb F)}^{-1}\mb F\cdot{\mb P}\big\|_2$ when the space-time kernel possesses uniform weighting form, then the result is generalized to general shape of kernel at zero temperature. In section \ref{sec:num_exp}, we carry out some numerical experiments of molecular dynamics to further understand our analysis results in the previous sections. In the final section, we draw some conclusions about our discoveries.

\section{Continuum stress in atomistic presentation}\label{sec:atom-to-cont}
\subsection{Continuum and atomistic system}
Continuum theory has been successfully used for decades to analyze and predict the mechanics of materials and structures. The core of the theory can be formulated as conservation laws, in the form
of mass, momentum and energy conservation. In continuum mechanics, these conservation
laws can be formulated in both Lagrangian (reference) coordinate or Eulerian (current)
coordinate, based on which the Piola-Kirchhoff stress or the Cauchy stress tensor could be derived. In this paper, our focus will be on the Eulerian reference and the Cauchy stress tensor.

We consider a system $\Omega$, its referential state (or initial state) is denoted as $\Omega_0$.
Let $\X\in\Omega_0\in\real^3$ be a point in the reference frame, $\x=\x(\X,t)\in\Omega\in\real^3$ be the point after deformation,
$\u(\x,t)=\x(\X,t)-\X$ be the displacement field and $\F=\frac{\partial \x(\X,t)}{\partial \X}$ be the deformation gradient. In Eulerian reference, continuum mechanics models are expressed as conservations of mass, momentum and energy:
\begin{subequations}
\begin{align}
\frac{\partial}{\partial t}\rho+\nabla_\x\cdot \q&=0,\label{eq:csv_1}\\
\frac{\partial}{\partial t}\q+\nabla_\x\cdot(\rho \v\otimes \v)&=\nabla_\x\cdot\bs\sigma,\label{eq:csv_2}\\
\frac{\partial}{\partial t}E+\nabla_\x\cdot(E\v+\J)&=\nabla_\x\cdot(\bs\sigma\cdot\v).\label{eq:csv_3}
\end{align}
\end{subequations}
In equation (\ref{eq:csv_1})-(\ref{eq:csv_3}), $\rho$ is mass density, $\q$ is momentum, $\v=\q/\rho$ is velocity, $\bs\sigma$ is Cauchy stress, $E$ is energy density, and $\J$ is heat flux.

It is worthwhile to notice that all the physical quantities defined above are under Eulerian reference. If we let $\mb{U}$ represent one of the above physical quantity, then $\mb{U}(\x,t)$ is uniquely determined by current location $\x$ and time t. Hence, all the derivative operator $\nabla_{\x}$ ($\nabla$ for simplicity) defined in this paper is spacial frame gradient rather than material frame gradient.

In atomistic model, we define $\X_i\in\Omega_0$ as reference position and $\x_i(t)\in\Omega$ as current position of the i-th atom, then the displacement of the i-th atom $\u_i(t)$ is written as $\u_i(t)=\x_i(t)-\X_i$. We assume that the interatomic potential $V$ can be uniquely determined by atom position $\{\x_i\}_{i=1}^N$. The motion of atoms are assumed to satisfy Newton's second Law:
\begin{equation}
m\ddot{\x}_i(t)=-\frac{\partial}{\partial \x_i}V(\x_1,\x_2,...,\x_n),
\end{equation}
where $m_i$ represents for mass of the i-th atom, $V(\x_i,...,\x_n)$ is the potential function determined by atoms position.
Let $\f_i$ be force exerting on the i-th atom, we assume there exists the force decomposition such that:
\begin{equation}
\f_i=\sum_{j\neq i}\f_{ij},\quad \f_{ij}=-\f_{ji}\label{eq:forc_dec},
\end{equation}
where $\f_{ij}$ is the force between the i-th atom and the $j$-th atom. One thing need to notice is that even if $V$ is not
pair-wise potential, there may still exists such force decomposition. To more detail, reader can see \cite{chen2006local,li2005multiscale}.

\subsection{Virial stress and Hardy stress}
\subsubsection{Virial stress}
For a homogeneous system, the average stress, referred to as the virial stress, has been widely used and studied. Here we introduce two versions of virial stress, the first of which is under Eulerian reference. Let $|\Omega|$ be the volume of the system, then the Eulerian version of virial stress is given by:
\begin{equation}
\label{eq:Virial_eulerian}
\widehat{\bs\sigma^{V}}(t)=
\left\langle\frac{-1}{2|\Omega|}\sum_{i=1}^{N}\sum_{i\neq j}\x_{ij}(t)\otimes{\f_{ij}(t)}
-\frac{1}{|\Omega|}\sum_{i=1}^{N}m_i\check{\v}_i(t)\otimes\check{\v}_i(t)\right\rangle_{(t)},
\end{equation}
where $\x_{ij}=\x_i-\x_j$, $\check{\v}_i(t) = \v_i(t)-\overline{\v}(t)$, $\v_i(t)$ is the velocity of $i$-th atom, $\overline{\v}(t)$ is the average velocity of all atoms in $\Omega$, and $\langle\cdot\rangle_{(t)}$ (or $\widehat{(\cdot)}$ ) represents for ensemble or temporal average.

The virial stress in Lagrangian reference is:
\begin{equation}\label{eq:Virial_lagrangian}
\widehat{{\mb P}^{V}}(t)=\left\langle\frac{-1}{2|\Omega_0|}\sum_{i=1}^{N}\sum_{i\neq j}\X_{ij}\otimes{\f_{ij}(t)}\right\rangle_{(t)},
\end{equation}
where $\X_{ij}=\X_i-\X_j$, $|\Omega_0|$ is volume of the system in referential state. In the following sections, $\bs\sigma^{V}$ and ${\mb P}^{V}$ will represent for the virial version of Cauchy and PK stress without temporal/ensemble average.

\subsubsection{Hardy stress}
For system out of equilibrium, or non-homogeneous system, one needs to compute the stress
locally at a point in space, the virial stress formula turns out to be inapplicable in this case \cite{cheung1991atomic}.
The Irving-Kirkwood formulism provides a rigorous approach to compute physical quantities in a local region, where they defined the empirical distribution as:
\begin{subequations}
\begin{align}
{\rho}(\x,t)&=\sum_{k=1}^{N}m_k\langle\delta(\x-\x_k),\rho(\x_1,...,\x_n,\q_1,...,\q_n;t)\rangle,\label{ik_def_1}\\
{\q}(\x,t)&=\sum_{k=1}^{N}\q_k\langle\delta(\x-\x_k),\rho(\x_1,...,\x_n,\q_1,...,\q_n;t)\rangle,\label{ik_def_2}
\end{align}
\end{subequations}
where $\delta(\x)$ is the Dirac Delta function, $\rho(\x_1,...,\x_n,\q_1,...,\q_n;t)$ is the density function of ensemble. $N$ is the number of the atoms in the system. The Cauchy stress tensor is derived based on Liouville equation and conservation equations in continuous mechanics:
\begin{equation}
\begin{aligned}
\bs\sigma^{I}(\x,t)=&-\frac{1}{2}\sum_{i=1}^{N}
    \sum_{j\neq i}\x_{ij}
    \otimes\f_{ij}\langle\int_0^1\delta\left(\x-\x_i+\lambda\x_{ij}\right)d\lambda,
    \rho(\x_1,...,\x_n,\q_1,...,\q_n;t)\rangle\\
    &-\sum_{k=1}^{N}m_k(\v_k-{\v})\otimes(\v_k-{\v})\langle\delta(\x-\x_k),\rho(\x_1,...,\x_n,\q_1,...,\q_n;t)\rangle.
\end{aligned}
\end{equation}

Hardy et al proposed another approach, which can be directly used in the MD simulation. Hardy's approach begins with representing the local quantities by using a kernel function $\varphi$:
\begin{subequations}
\begin{align}
{\rho}(\x,t)&=\sum_{k=1}^{N}m_k\varphi(\x-\x_k(t)),\label{eq:hardy_def_1}\\
{\q}(\x,t)&=\sum_{k=1}^{N}m_k \v_k(t)\varphi(\x-\x_k(t)),\label{eq:hardy_def_2}\\
{e}(\x,t)&=\frac{1}{2}\sum_{k=1}^{N}\left(m_k {|\v_k(t)|}^2+V_k(t)\right)\varphi(\x-\x_k(t)).\label{eq:hardy_def_3}
\end{align}
\end{subequations}
In equation (\ref{eq:hardy_def_1})-(\ref{eq:hardy_def_3}), $\varphi$ can be considered as a weight function, taking average of physical quantities near sample point $\x$. In connection to the IK's formulism, the function $\varphi$ can be viewed as a regularization to the Dirac Delta function. It can be also considered as the average of the function $\varphi(\mb x-\mb y)$ with respect to the one-particle distributions.

Hardy's criteria \cite{hardy1982formulas} will be taken as guidelines to select these $\varphi$:
\begin{itemize}
\item
$\varphi(\x)$ has maximum at $\x=0$.
\item
$\varphi(\x)\rightarrow0\ as\ |\x|\rightarrow+\infty$.
\item
$\varphi(\x)$ is smooth and non-negative.
\item
$\int_{\real^3}\varphi(\x)dx=1$.
\end{itemize}

For the integrity of the paper, we briefly explain how the stress can be derived from this approach, the derivations could be found in lots of materials including \cite{hardy1982formulas,murdoch1994continuum,zimmerman2004calculation,webb2008reconsideration}.
From the equations (\ref{eq:hardy_def_2}), one gets
\begin{equation*}
\begin{aligned}
\frac{\partial}{\partial t}\q(\x,t)
=&\frac{1}{2}\sum_{i=1}^{N}\sum_{j\neq i}\f_{ij}(t)\left(\varphi(\x-\x_i(t))-\varphi(\x-\x_j(t))\right)\\
    &-\nabla\cdot\left(\sum_{k=1}^{N}m_k\v_k(t)\otimes\v_k(t)\varphi(\x-\x_k(t))\right),
\end{aligned}
\end{equation*}
where (\ref{eq:forc_dec}) is used to achieve force decomposition.

At this point, a ``bond function" $b_{ij}(\x,t)$ is defined as:
\begin{equation}\label{eq:hardy_bond}
b_{ij}(\x,t)=\int_{0}^{1}\varphi(\x-\x_i(t)+\lambda \x_{ij}(t))d\lambda.
\end{equation}
Then we can write the expression as:
\begin{equation*}
\begin{aligned}
\frac{\partial}{\partial t}\q(\x,t)
=&-\frac{1}{2}\sum_{i=1}^{N}\sum_{j\neq i}\f_{ij}\x_{ij}\cdot\partial_X b_{ij}(\x,t)\\
    &-\nabla\cdot\left(\sum_{k=1}^{N}\varphi(\x-\x_k(t))m_k\v_k(t)\otimes\v_k(t)\right).
\end{aligned}
\end{equation*}
Here, we introduce the ``revised velocity"
\begin{equation}\label{eq:hardy_rv}
\mb w_i(\x,t)=\v_i(t)-\v(\x,t),
\end{equation}
then there holds:
\begin{equation*}
\begin{aligned}
\frac{\partial}{\partial t}\q(\x,t)
=&\nabla\cdot\left(-\frac{1}{2}\sum_{i=1}^{N}\sum_{j\neq i}\x_{ij}(t)\otimes\f_{ij}(t)b_{ij}(\x,t)\right)\\
    &-\nabla\cdot\left(\sum_{k=1}^{N}m_k\mb w_k(\x,t)\otimes\mb w_k(\x,t)\varphi(\x-\x_k(t))\right)
    -\nabla\cdot(\rho\v\otimes\v).
\end{aligned}
\end{equation*}
Combined with equation (\ref{eq:csv_2}), we get the Hardy stress
\begin{equation}\label{eq:hardy_cauchy_stress}
\begin{aligned}
\bs\sigma^H(\x,t)=&-\frac{1}{2}\sum_{i=1}^{N}\sum_{j\neq i}\x_{ij}(t)\otimes\f_{ij}(t)b_{ij}(\x,t)\\[3pt]
    &-\sum_{k=1}^{N}m_k\mb w_k(\x,t)\otimes\mb w_k(\x,t)\varphi(\x-\x_k(t)).
\end{aligned}
\end{equation}

\section{Generalized Irving-Kirkwood formulism}\label{sec:generalized-IK}
\subsection{Definitions of mass density and momentum in generalized IK formulism}\label{sec:kernel_list}
In molecular dynamics simulation, the direct usage of Hardy stress will cause irrational fluctuations even in a homogeneous system. This is caused by the inconsistency of time-scales between the atomistic and continuum systems.
Therefore, it needs to be further processed to obtain a more stable value through time or ensemble average. In the original IK formalism, the ensemble average was used. Here
we propose to extend the IK formulism to naturally incorporate time averaging by using a space-time kernel $\Phi(\x,t)$.

Our former article \cite{yang2012generalized} surveys on g-IK formulism in Lagrangian reference, where the momentum and mass density are defined as
\begin{equation*}
\begin{aligned}
\tilde{\rho^L}(\mb{X},t)&=\sum_{i=1}^N m_i\int_\mathbb{R}\Phi(\mb{X}-\mb{X}_i,t-s)ds,\\
\tilde{\mb{q}^L}(\mb{X},t)&=\sum_{i=1}^N m_i\int_\mathbb{R}\Phi(\mb{X}-\mb{X}_i,t-s)\mb{v}_i(s)ds,
\end{aligned}
\end{equation*}
where $\Phi(\X,t)$ is the space-time kernel function coupling atomistic quantities ($m_i$ and $\v_i$) with
continuum quantities ($\tilde{\rho^L}$ and $\tilde{\mb{q}^L}$).
A natural way to transport the definitions to Eulerian reference may be
\begin{equation*}
\begin{aligned}
\tilde{\rho'}(\mb{x},t)&=\sum_{i=1}^N m_i\int_\mathbb{R}\Phi(\mb{x}-\mb{x}_i(t),t-s)ds,\\
\tilde{\mb{q'}}(\mb{x},t)&=\sum_{i=1}^N m_i\int_\mathbb{R}\Phi(\mb{x}-\mb{x}_i(t),t-s)\mb{v}_i(s)ds.
\end{aligned}
\end{equation*}
However, these definitions turn out to violate the conservation of mass. For we have
\begin{equation*}
\begin{aligned}
\frac{\partial}{\partial t}\tilde\rho'(\x,t)
=&\frac{\partial}{\partial t}\sum_{i=1}^N m_i\int_\mathbb{R}\Phi(\mb{x}-\mb{x}_i(t),t-s)ds\\
=&\nabla_\x\cdot\left(\sum_{i=1}^N m_i\int_\mathbb{R}\Phi(\mb{x}-\mb{x}_i(t),t-s)(-\v_i(t))ds\right).\\
\end{aligned}
\end{equation*}
Thus
\begin{equation*}
\begin{aligned}
\frac{\partial}{\partial t}\tilde\rho'(\x,t)+\nabla_\x\cdot\tilde\q'(\x,t)
=&\nabla_\x\cdot\left(\sum_{i=1}^N m_i\int_\mathbb{R}\Phi(\mb{x}-\mb{x}_i(t),t-s)(\v_i(s)-\v_i(t))ds\right)\neq0.
\end{aligned}
\end{equation*}
So, some care must be paid in the definitions of mass density and momentum in the g-IK formulism when we consider them in Eulerian reference.

Another possible way to define mass density and momentum is
\begin{subequations}\label{eq:IK-new}
\begin{align}
\tilde{\rho}(\mb{x},t)&=\sum_{i=1}^N m_i\int_\mathbb{R}\Phi(\mb{x}-\mb{x}_i(s),t-s)ds,\\
\tilde{\mb{q}}(\mb{x},t)&=\sum_{i=1}^N m_i\int_\mathbb{R}\Phi(\mb{x}-\mb{x}_i(s),t-s)\mb{v}_i(s)ds.
\end{align}
\end{subequations}
Here, we again first check whether these definitions are consistent with mass conservation. By taking derivative of $\tilde\rho$, there is
\begin{equation*}
\begin{aligned}
\frac{\partial}{\partial t}\tilde\rho(\x,t)
=&\frac{\partial}{\partial t}\sum_{i=1}^N m_i\int_\mathbb{R}\Phi(\mb{x}-\mb{x}_i(t-\xi),\xi)d\xi\\
=&\nabla_\x\cdot\left(\sum_{i=1}^N m_i\int_\mathbb{R}\Phi(\mb{x}-\mb{x}_i(t-\xi),\xi)(-\v_i(t-\xi))d\xi\right)\\
=&-\nabla_\x\cdot\tilde\q(\x,t).
\end{aligned}
\end{equation*}
Thus, there holds
\begin{equation*}
\frac{\partial}{\partial t}\tilde\rho(\x,t)+\nabla_\x\cdot\tilde\q(\x,t)=0,
\end{equation*}
namely, the definitions of $\tilde\rho$ and $\tilde\q$ (\ref{eq:IK-new}) are consistent with mass conservation equation (\ref{eq:csv_1}). Naturally, these definitions will be taken as the ones to derive the Cauchy stress tensor in the next section.

Now, we discuss the several properties of the space-time kernel $\Phi$. We bare in mind the Hardy's criteria of choosing the spacial kernel function and try to generalize these criteria to the 4 dimensional space-time.
It is often convenient to start with an original kernel $\Psi$ which has a support in an unit cubic, and then define
\begin{equation*}
\Phi(\x,t)=\frac{1}{r_s^3r_t}\Psi(\x/r_s,t/r_t),
\end{equation*}
where $r_s$ is spacial radius, and $r_t$ is temporal radius of $\Phi$. For $\Psi$, we give several criteria similar to Hardy's:
\begin{itemize}
\item
$\forall t\in\real$, $\Psi(\x,t)$ has maximum at $\x=0$.
\item
$\{(\x,t)\in\real^4:\Psi(\x,t)\neq0\}\subseteq[-1,1]^3\times[0,1]$.
\item
$\Psi(\x,t)\in\mathcal C^1\left([-1,1]^3\times[0,1]\right)$.
\item
$\int_{\real^4}\Psi(\x,t)dx=1$.
\end{itemize}
Here, we give several choices for $\Psi$ that satisfy the above criteria:
\begin{enumerate}
\item
Type I:
\begin{equation*}
\Psi^{\textrm{I}}(x,y,z,t)
=\left\{\begin{array}{rl}
\dps\frac{1}{8}\prod_{\xi=x,y,z,t}(1+\cos(\pi\xi)),&|x_1|\le 1,\ |x_2|\le 1,\ |x_3|\le 1,\ 0\le{t}\le 1,\\
0,&{\rm Otherwise}.
\end{array}\right.
\end{equation*}
\item
Type II:
\begin{equation*}
\Psi^{\textrm{II}}(x_1,x_2,x_3,t)=
\left\{\begin{array}{rl}
1/8,&|x_1|\le 1,\ |x_2|\le 1,\ |x_3|\le 1,\ 0\le{t}\le 1,\\[5pt]
0,&\text{Otherwise}.
\end{array}\right.
\end{equation*}
\item
Type III:
\begin{equation*}
\Psi^{\textrm{III}}(x,y,z,t)=
\left\{\begin{array}{rl}
\displaystyle\frac{15}{4\pi}\left[1+\left(2{r}-3\right)r^2\right],&r\le 1\;\text{and}\;1\ge t\ge0,\\[5pt]
0,&\text{Otherwise}.
\end{array}\right.
\end{equation*}
Here \(r=\sqrt{x^2+y^2+z^2}.\)
\item
Type IV:
\begin{equation*}
\Psi^{\textrm{IV}}(x,y,z,t)=
\left\{\begin{array}{rl}
\displaystyle\frac{7}{\pi^2}\left[1+\left(2{R}-3\right)R^2\right],&R\le 1\;\text{and}\;t\ge0,\\[5pt]
0,&\text{Otherwise}.
\end{array}\right.
\end{equation*}
Here \(R=\sqrt{x^2+y^2+z^2+t^2}.\)
\end{enumerate}

\subsection{Derivation for Cauchy stress in generalized Irving-Kirkwood formulism}
In this subsection, we try to derive the Cauchy stress in g-IK formulism. We start from definition (\ref{eq:IK-new}),
by taking time derivative of $\tilde{\mb q}$, there is
\begin{equation*}
  \begin{aligned}
    \frac{\partial\tilde{\mb{q}}(\mb{x},t)}{\partial t}
    =&\frac{\partial}{\partial t}\sum_{i=1}^Nm_i\int_\real\Phi(\mb{x}-\mb{x}_i(s),t-s)\mb{v}_i(s)ds\\
    \overset{\xi=t-s}{=\!=\!=\!=}&\sum_{i=1}^N\int_\real \Phi(\mb{x}-\mb{x}_i(t-\xi),\xi)\mb{f}_i(t-\xi)d\xi\\
    &-\sum_{i=1}^Nm_i\int_\real \mb{v}_i(t-\xi)\partial_{\mb{x}}\Phi(\mb{x}-\mb{x}_i(t-\xi),\xi)\cdot \mb{v}_i(t-\xi)d\xi\\
    \overset{s=t-\xi}{=\!=\!=\!=}&\sum_{i=1}^N\int_\real \Phi(\mb{x}-\mb{x}_i(s),t-s)\sum_{j\ne i}\mb{f}_{ij}(s)ds\\
    &-\sum_{i=1}^Nm_i\int_\real \mb{v}_i(s)\partial_{\mb{x}}\Phi(\mb{x}-\mb{x}_i(s),t-s)\cdot \mb{v}_i(s)ds.\\
    =&\int_\real\frac{1}{2}\sum_{i=1}^N\sum_{j\ne i}\mb{f}_{ij}(s)(\Phi(\mb{x}-\mb{x}_i(s),t-s)
    -\Phi(\mb{x}-\mb{x}_j(s),t-s))ds\\
    &-\sum_{i=1}^Nm_i\int_\real \mb{v}_i(s)\partial_{\mb{x}}\Phi(\mb{x}-\mb{x}_i(s),t-s)\cdot \mb{v}_i(s)ds.
    \end{aligned}
\end{equation*}
Here, we define the generalized ``bond function" $B_{ij}(\mb x,s,t-s)$ as
\begin{equation}\label{eq:gik_bond}
  B_{ij}(\mb x,s,t-s)= \int_0^1\Phi\left(\mb{x}-(\mb{x}_i(s)+\lambda \mb{x}_{ji}(s)), t-s\right)d\lambda.
\end{equation}
Then there is
\begin{equation*}
\begin{aligned}
\frac{\partial \tilde {\mb{q}}(\mb{x},t)}{\partial t}
    =&\nabla_{\mb{x}}\cdot\left(-\frac{1}{2}\sum_{i=1}^N\sum_{j\neq i}\int
    B_{ij}(\mb x,s,t-s)\mb{x}_{ij}(s)\otimes\mb{f}_{ij}(s)ds\right)\\
    &\nabla_{\mb{x}}\cdot\left(-\sum_{i=1}^N m_i\int \mb{v}_i(s)\otimes
    \mb{v}_i(s)\Phi(\mb x-\mb x_i(s), t-s)ds\right).
\end{aligned}
\end{equation*}
Now, we define the generalized ``revised velocity" $\tilde{\w}_i(\x,s,t)$:
\begin{equation}\label{eq:gik_rv}
\tilde{\w}_i(\x,s,t)=\v_i(s)-\tilde\v(\x,t),
\end{equation}
then there hold
\begin{equation*}
\begin{aligned}
&\sum_{i=1}^N m_i\int\tilde{\w}_i(\x,s,t)\otimes
    \tilde{\w}_i(\x,s,t)\Phi(\mb x-\mb x_i(s), t-s)ds\\
&=\sum_{i=1}^N m_i\int{\v}_i(s)\otimes
    {\v}_i(s)\Phi(\mb x-\mb x_i(s), t-s)ds\\
&-\sum_{i=1}^N m_i\int\tilde\v(\x,t)\otimes
    {\v}_i(s)\Phi(\mb x-\mb x_i(s), t-s)ds\\
&-\sum_{i=1}^N m_i\int{\v}_i(s)\otimes
    \tilde\v(\x,t)\Phi(\mb x-\mb x_i(s), t-s)ds\\
&+\sum_{i=1}^N m_i\int\tilde\v(\x,t)\otimes
    \tilde\v(\x,t)\Phi(\mb x-\mb x_i(s), t-s)ds\\
&=\sum_{i=1}^N m_i\int{\v}_i(s)\otimes
    {\v}_i(s)\Phi(\mb x-\mb x_i(s), t-s)ds
    -2(\tilde\q\otimes\tilde\v)+(\tilde\q\otimes\tilde\v)
\end{aligned}
\end{equation*}
Hence, if we define the Cauchy stress $\bs\sigma^G$ as
\begin{equation}\label{eq:new_ik_stress}
  \begin{aligned}
     {\bs{\sigma}^{G}}(\mb{x},t)=
    &-\frac{1}{2}\sum_{i=1}^N\sum_{j\neq i}\int B_{ij}(\mb x,s,t-s)\mb{f}_{ij}(s)\otimes \mb{x}_{ij}(s)ds\\
    &-\sum_{i=1}^N m_i\int\tilde{\mb{w}}_i(\x,s,t)\otimes\tilde{\mb{w}}_i(\x,s,t)\Phi(\mb{x}-\mb{x}_i(s),t-s)ds.
  \end{aligned}
\end{equation}
It satisfies the momentum conservation law (\ref{eq:csv_2}):
\begin{equation*}
\frac{\partial}{\partial t}\tilde\q+\nabla_\x\cdot(\tilde\q\otimes\tilde\v)=\nabla_\x\cdot\bs\sigma^G.
\end{equation*}
Thus, \eqref{eq:new_ik_stress} will be our definition of g-IK stress. Obviously, $\tilde\q$, $\tilde\rho$ and $\bs\sigma^G$ are consistent with conservation laws.

\section{Comparison between generalized IK formulism and traditional temporal post-process approach}\label{sec:g-IK-t-tpp}
In traditional temporal post-process approach, there are basically two ways to derive the stress tensor:
\begin{enumerate}
\item
\textbf{Time-averaged stress}: In this approach, the Hardy stress is first derived, then it is averaged with a temporal kernel function.
\item
\textbf{Murdoch's stress} \cite{murdoch1994continuum}: In this approach, the mass density and momentum defined in Hardy's approach are first averaged with a temporal kernel function, then a revised version of stress tensor is derived based on the temporally averaged mass density and momentum.
\end{enumerate}
A common mark of the two ways of temporal post-process approach is, the spacial and temporal average are two separable steps. This mark of temporal post-process approach will lead to the first difference with g-IK approach, namely, the inseparable kernel case. We say the kernel $\Phi(\x,t)$ is space-time separable, when it has the following decomposition
\begin{equation}\label{eq:separable}
\Phi(\x,t)=\varphi(\x)\tau(t),
\end{equation}
where $\varphi(\x)$ and $\tau(t)$ could be regarded as spacial kernel and temporal kernel respectively. Otherwise, we say the kernel $\Phi(\x,t)$ is inseparable.

\begin{figure}[htbp]
\begin{center}
\includegraphics[scale=0.5]{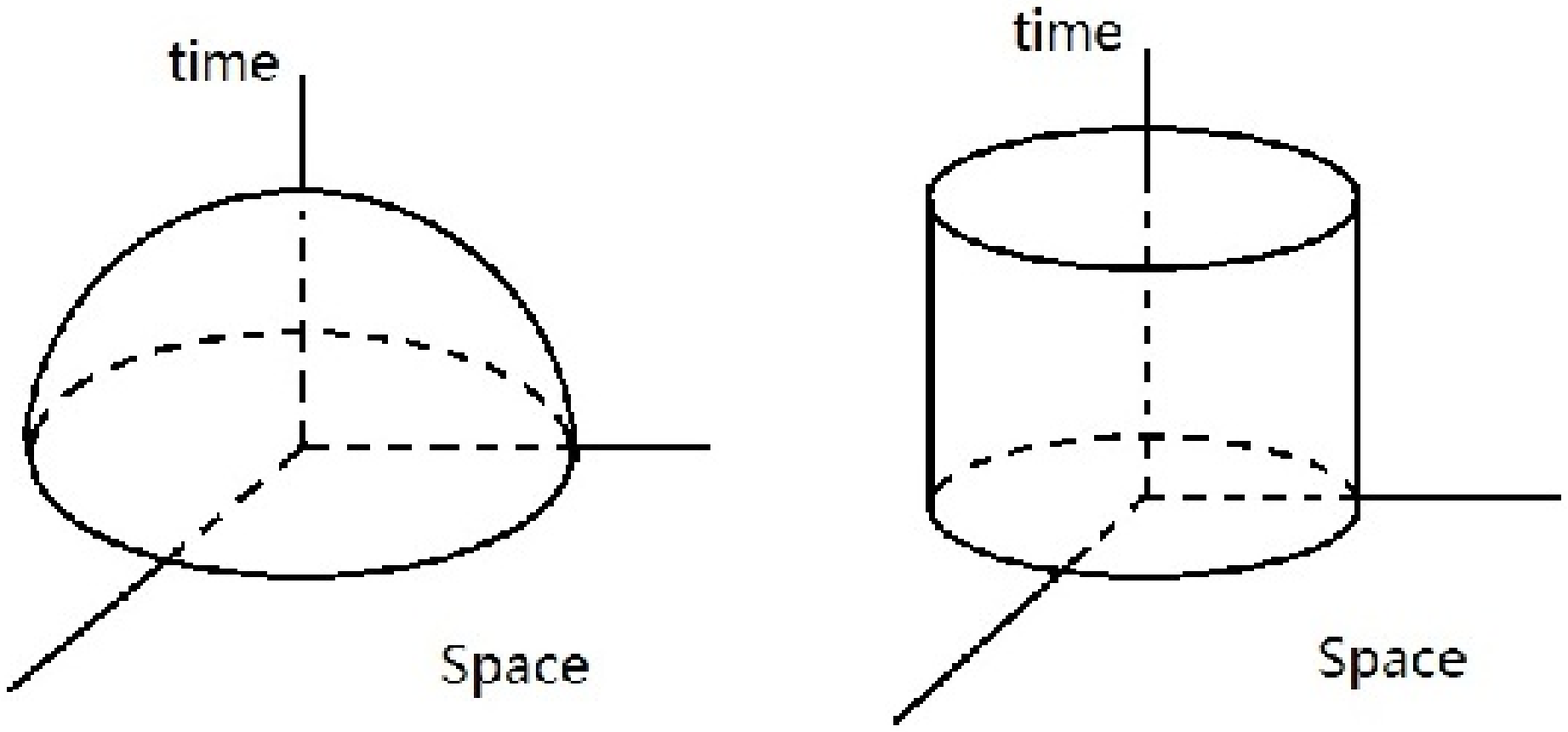}
\end{center}
\caption{The left and the right figure respectfully show the support domain of $\Psi^{\textrm{IV}}$ and $\Psi^{\textrm{III}}$.}
\label{fig:kernel}
\end{figure}

An inseparable kernel will make the steps of spacial and temporal average into an indistinguishable unified process, in which case the separation of the two steps would be impossible. One feature of an inseparable space-time kernel is, the spacial average domain will change as time. The left figure in Figure \ref{fig:kernel} shows the support domain of a typical inseparable kernel (the Type IV kernel in section \ref{sec:kernel_list}). We can see, when the averaging time is near the sample time (the center of the space-time), the spacial average domain increases. This feature gives us more choice of coupling continuum mechanics quantities with atomistic information, as it allows us to give different domain of spacial average at different time, while for the separable kernel (for instance, the Type III kernel, right figure of Figure \ref{fig:kernel}), the spacial average domain is always the same.

Another thing worth our consideration is, when the kernel function is space-time separable, whether the g-IK stress can be reduced to one of the above two versions of stress defined in temporal post-process approach. To figure out this problem, we will now assume our following discussions are in the sense of a separable kernel (otherwise, the discussion is meaningless as we shown above).
In \cite{yang2012generalized}, we show that the above two versions of stress in temporal post-process approach and g-IK stress are all identical in Lagrangian reference.
Here, in Eulerian reference, our discoveries are
\begin{enumerate}
\item
The g-IK stress could be regarded as Murdoch's stress.
\item
There lies difference between g-IK stress and time-averaged stress, the value of the difference is related to the spacial and temporal radius of the kernel function.
\end{enumerate}
In the next two subsections, we will respectively study these two cases.

\subsection{Consistency between generalized IK formulism and traditional temporal post-process approach}
In this part, we will show that Murdoch's stress could be regarded as the g-IK stress, when the kernel $\Phi(\x,t)$ is space-time separable. In this case, the definitions for mass density and momentum (\ref{eq:IK-new}) become
\begin{subequations}
\begin{align}
\tilde{\rho}(\mb{x},t)&=\int_\mathbb{R}\tau(t-s)\sum_{i=1}^N m_i\varphi(\mb{x}-\mb{x}_i(s))ds=\hat\rho,\\
\tilde{\mb{q}}(\mb{x},t)&=\int_\mathbb{R}\tau(t-s)\sum_{i=1}^N m_i\v_i(s)\varphi(\mb{x}-\mb{x}_i(s))ds=\hat\q,
\end{align}
\end{subequations}
where $\hat\rho:=\rho*\tau$, $\hat\q:=\q*\tau$. We notice that, now the mass density and momentum defined in g-IK formulism are equal to the ones by making temporal post-process on Hardy's definitions for mass density and momentum with a temporal kernel function $\tau$.
Starting from $\hat\rho$ and $\hat\q$, Murdoch \cite{murdoch1994continuum} derived the Cauchy stress tensor as
\begin{equation*}
\overline\T_u:=\overline{\T^-}-\overline\D_u,
\end{equation*}
where
\begin{equation*}
\overline{\T^-}=-\frac{1}{2}\int\tau(t-s)ds\big(\sum_{i=1}^N\sum_{j\neq i}b_{ij}(\mb x,s)\mb{x}_{ij}(s)\otimes\mb{f}_{ij}(s)\big),
\end{equation*}
and
\begin{equation*}
\overline\D_u=\int\tau(t-s)ds\big(\sum_{i=1}^N m_i\tilde{\mb{w}}_i(\x,s,t)\otimes\tilde{\mb{w}}_i(\x,s,t)\varphi(\mb{x}-\mb{x}_i(s))\big).
\end{equation*}
We notice that when $\Phi$ is separable, the ``bond function" in Hardy's (\ref{eq:hardy_bond}) and g-IK's (\ref{eq:gik_bond}) approaches have the connection:
\begin{equation*}
B_{ij}(\mb x,s,t-s)=\tau(t-s)b_{ij}(\x,s),
\end{equation*}
thus it is obvious that
\begin{equation*}
\overline\T_u=\bs\sigma^G.
\end{equation*}
So, the g-IK formulism is consistent with the traditional temporal post-process approach.

\subsection{Difference between generalized IK stress and time-averaged Hardy stress}
In this part, we will compare the g-IK stress with time-averaged Hardy stress in the case of a separable kernel. When we look at the expressions of the g-IK stress (or equivalently, the Murdoch's definition of stress $\overline\T_u$) and the time-averaged Hardy stress, the only difference between the two lies in the kinetic part of the stress. More precisely, the ``revised velocity" of the two versions of stress ((\ref{eq:hardy_rv}) and (\ref{eq:gik_rv})) are different, for we have
\begin{equation*}
\tilde{\w}_i(\x,s,t)=\v_i(s)-\tilde\v(\x,t),\ \mb w_i(\x,s)=\v_i(s)-\v(\x,s),
\end{equation*}
where $\tilde\w_i$ is obtained in g-IK formulism, $\mb w_i$ is obtained in Hardy's approach. The physical meaning of the revised velocity can be interpreted as the particle oscillations relative to the macroscopic field. In g-IK formulism, the macroscopic velocity field is naturally incorporated with spacial and temporal average, but in Hardy's approach, the field is only incorporated with spacial average. Thus a direct temporal average on the Hardy's stress may be unreasonable as the macroscopic velocity field is only up to a sense of spacial average while the stress is defined under the meaning of space-time average.

Another approach to understand the difference is to first consider again the momentum conservation equation (\ref{eq:csv_2}), if we take temporal average at both sides of the equation, there hold
\begin{equation*}
\frac{\partial}{\partial t}\hat\q+\nabla_\x\cdot(\widehat{\q\otimes\v})=\nabla_\x\cdot\widehat{\bs\sigma^H}.
\end{equation*}
On the other hand, there is
\begin{equation*}
\frac{\partial}{\partial t}\hat\q+\nabla_\x\cdot({\hat\q\otimes\frac{\hat\q}{\hat\rho}})=\nabla_\x\cdot{\bs\sigma^G}.
\end{equation*}
Thus
\begin{equation}\label{eq:moment_csv_g-i-k_t_pp}
\nabla_\x\cdot({\bs\sigma^G}-\widehat{\bs\sigma^H})=\nabla_\x\cdot({\hat\q\otimes\frac{\hat\q}{\hat\rho}}-\widehat{\q\otimes\v}).
\end{equation}
As $\tilde\q$, $\tilde\rho$ and $\bs\sigma^G$ satisfy the conservation of momentum, and generally, $\nabla_\x\cdot({\hat\q\otimes\frac{\hat\q}{\hat\rho}}-\widehat{\q\otimes\v})\neq0$. So $\tilde\q$, $\tilde\rho$ and $\widehat{\bs\sigma^H}$ do not satisfy conservation of momentum in general.

Now we try to figure out the difference between ${\bs\sigma^G}$ and $\widehat{\bs\sigma^H}$ directly, according to equation (\ref{eq:new_ik_stress}) and (\ref{eq:hardy_cauchy_stress}), we have
\begin{equation*}
\begin{aligned}
    &\widehat{\bs\sigma^{H}}(\x,t)-\bs\sigma^{G}(\x,t)\\
    =&\sum_{i=1}^N m_i\int\varphi(\x-\x_i(s))\tau(t-s)
        \left[\w_i(\x,s)\otimes\w_i(\x,s)-\tilde\w_i(\x,s,t)\otimes\tilde\w_i(\x,s,t)\right]ds\\
    =&\sum_{i=1}^N m_i\int\varphi(\x-\x_i(s))\tau(t-s)
        \left[\w_i(\x,s)\otimes(\v(\x,s)-\tilde\v(\x,t))\right]ds\\
    &+\sum_{i=1}^N m_i\int\varphi(\x-\x_i(s))\tau(t-s)
        \left[(\v(\x,s)-\tilde\v(\x,t))\otimes\tilde\w_i(\x,s,t)\right]ds\\
    =&\int\tau(t-s)\left\{\sum_{i=1}^N m_i\varphi(\x-\x_i(s))
        \w_i(\x,s)\otimes[\v(\x,s)-\tilde\v(\x,t)]\right\}ds\\
    &+\int\tau(t-s)\left\{[\v(\x,s)-\tilde\v(\x,t)]\otimes
        \sum_{i=1}^N m_i\varphi(\x-\x_i(s))\tilde\w_i(\x,s,t)\right\}ds.
\end{aligned}
\end{equation*}
Here we notice the fact that
\begin{equation*}
\begin{aligned}
    \sum_{i=1}^N m_i\varphi(\x-\x_i(s))\w_i(\x,s)=0,\\
    \int\tau(t-s)\left\{\sum_{i=1}^N m_i\varphi(\x-\x_i(s))\tilde\w_i(\x,s,t)\right\}ds=0.
\end{aligned}
\end{equation*}
So there hold
\begin{equation*}
\begin{aligned}
    &\widehat{\bs\sigma^{H}}(\x,t)-\bs\sigma^{G}(\x,t)\\
    =&\int\tau(t-s)\left\{\v(\x,s)\otimes
        \sum_{i=1}^N m_i\varphi(\x-\x_i(s))(\v_i(s)-\tilde\v(\x,t))\right\}ds\\
    =&\int\tau(t-s)\left\{\v(\x,s)\otimes
        [\q(\x,s)-\rho(\x,s)\tilde\v(\x,t)]\right\}ds\\
    =&\int\tau(t-s)\left\{\q(\x,s)\otimes
        [\v(\x,s)-\tilde\v(\x,t)]\right\}ds.\\
\end{aligned}
\end{equation*}
Now we use the fact $\displaystyle\tilde\v=\frac{\tilde\q}{\tilde\rho}=\frac{\hat\q}{\hat\rho}$, and get
\begin{equation}\label{eq:diff_g-i-k_t_pp}
\widehat{\bs\sigma^{H}}(\x,t)-\bs\sigma^{G}(\x,t)=\widehat{\q\otimes\frac{\q}{\rho}}-{\hat\q\otimes\frac{\hat\q}{\hat\rho}}.
\end{equation}
Obviously, this is consistent with equation (\ref{eq:moment_csv_g-i-k_t_pp}). But we can not directly conclude (\ref{eq:diff_g-i-k_t_pp}) from (\ref{eq:moment_csv_g-i-k_t_pp}), as the stress tensor is unique only up to a divergence free term.

Murdoch \cite{murdoch1993physical} has investigated the right term of \eqref{eq:diff_g-i-k_t_pp}, and reasoned that it can not be neglected when $\v^I-\tilde\v$ varies significantly in space
and/or time at the $r_s$ (spacial radius), $r_t$ (temporal radius) scales, where $\v^I$ could be regarded as the velocity defined in the original IK formulism. Here we try to investigate the connection between the value of the right term of \eqref{eq:diff_g-i-k_t_pp} and the spacial/temporal radius $r_s/r_t$. From our formulas above, we know
\begin{equation}\label{eq:gik_tpp_diff}
\begin{aligned}
\|\widehat{\bs\sigma^{H}}(\x,t)-\bs\sigma^{G}(\x,t)\|_2
    \le\int\tau(t-s)\|\q(\x,s)\|_2\|\v(\x,s)-\tilde\v(\x,t)\|_2ds.
\end{aligned}
\end{equation}
So the difference is controlled by the mean value of $\|\v(\x,s)-\tilde\v(\x,t)\|_2$. Intuitively, there are two situations in which the difference maybe small, we will respectively check these two situations. The first one is when the temporal radius $r_t$ goes down to zero. In this situation, we have
\begin{equation*}
\tau\rightarrow\delta,
\end{equation*}
where $\delta$ represents for the Dirac Delta function. As Dirac Delta function is the identity of convolution group,
there hold
\begin{equation*}
\begin{aligned}
&\tilde\v=\frac{\hat\q(\x,t)}{\hat\rho(\x,t)}
=\frac{\q*\tau}{\rho*\tau}\rightarrow\frac{\q*\delta}{\rho*\delta}=\frac{\q}{\rho}=\v,\\
&\tau*\|\q(\x,s)\|_2\|\v(\x,s)-\tilde\v(\x,t)\|_2\\
&\rightarrow\delta*\|\q(\x,s)\|_2\|\v(\x,s)-\tilde\v(\x,t)\|_2=\|\q(\x,t)\|_2\|\v(\x,t)-\tilde\v(\x,t)\|_2.
\end{aligned}
\end{equation*}
So we know the difference goes down to zero as the temporal radius goes down to zero.

Another situation is when the spacial average radius ($r_s$) becomes large enough. In this situation, the physical quantities obtained from spacial average become more stable to time. This could be interpreted as
\begin{equation}\label{eq:stable_assump}
  (\|(\f*\tau)(t)-\f\|_2*\tau)(t)\rightarrow0,\ {\rm as}\ r_s>>1,
\end{equation}
where $\f$ is some spatially averaged macroscopic local field.
When $\tau(x)=\frac{1}{r_t}\chi_{[x-r_t,x]}$, the meaning of the above equation becomes clear, it is just the variance of $\f$ (in the sense of absolute value) goes down to zero. Now we consider again the equation \eqref{eq:gik_tpp_diff} and have
\begin{equation*}
\begin{aligned}
&\|\widehat{\bs\sigma^{H}}(\x,t)-\bs\sigma^{G}(\x,t)\|_2
    \le(\tau*[\|\q\|_2\|\v-\tilde\v(t)\|_2])(t)\\
    &=(\tau*[\|\q\|_2\left\|\frac{\q}{\rho}-\frac{(\q*\tau)(t)}{(\rho*\tau)(t)}\right\|_2])(t)\\
    &=(\tau*[\|\q\|_2\left\|\frac{\q(\rho*\tau)(t)-\rho(\q*\tau)(t)}{\rho(\rho*\tau)(t)}\right\|_2])(t)\\
    &=(\tau*[\|\q\|_2\left\|\frac{\q[(\rho*\tau)(t)-\rho]-\rho[(\q*\tau)(t)-\q]}{\rho(\rho*\tau)(t)}\right\|_2])(t)\\
    &\le\sup_{s}[\frac{\|\q(s)\|_2^2}{\rho(s)(\rho*\tau)(t)}](\tau*\|(\rho*\tau)(t)-\rho\|_2)(t)\\
    &+\sup_s[\frac{\|\q(s)\|_2}{(\rho*\tau)(t)}](\tau*\|(\q*\tau)(t)-\q\|_2)(t).
\end{aligned}
\end{equation*}
According to \eqref{eq:stable_assump}, there hold
\begin{equation*}
(\tau*\|(\q*\tau)(t)-\q\|_2)(t)\rightarrow0,\ (\tau*\|(\rho*\tau)(t)-\rho\|_2)(t)\rightarrow0,
\end{equation*}
so the difference goes down as the spacial radius $r_s$ increase.
The numerical experiments
presented in section \ref{sec:num_exp}, suggest that the difference between g-IK stress and time-averaged stress decreases when $r_s$ increases or $r_t$ decreases to zero as we show in this section.

\section{Reconsideration of several versions of stress in generalized IK formulism}\label{sec:pk-cauchy-in-gIK}
\subsection{Generalized IK stress and virial stress}
The virial stress tensor can be re-derived from the time-averaged Hardy stress at a special case of a uniform weighting spacial kernel function \cite{admal2010unified}. In this subsection, we will study the connection between virial stress and g-IK stress at the case when the space-time kernel has a uniform weighting form. As the g-IK stress naturally incorporate temporal average, the connection can be directly build. However, some care must be paid on the difference between the g-IK stress and time-averaged stress when building the connection.

To begin with,
we say the space-time kernel function $\Phi$ has the uniform weighting form, when it can be written as
\begin{equation}\label{eq:step_kernel}
\Phi(x_1,x_2,x_3,t)=
\left\{\begin{array}{rl}
1/(8r_s^3r_t),&|x_1|\le r_s,\ |x_2|\le r_s,\ |x_3|\le r_s,\ 0\le{t}\le r_t,\\[5pt]
0,&{\rm else},
\end{array}\right.
\end{equation}
where $r_s$ is spacial radius, $r_t$ is temporal radius. One thing need to notice is that $\Phi$ is separable in this case, namely
\begin{equation*}
\begin{aligned}
\Phi(\x,t) &= \varphi(\x)\tau(t),\\
\varphi(\x) &= 1/(8r_s^3),\ |x_1|\le r_s,\ |x_2|\le r_s,\ |x_3|\le r_s,\\
\tau(t) &= 1/(r_t),\ 0\le{t}\le r_t.
\end{aligned}
\end{equation*}
In this case, we have
\begin{equation*}
\bs\sigma^H = \bs\sigma^V - \sum_{\x_{ij}\cap\partial\Omega}\x_{ij}\otimes\f_{ij}b_{ij}.
\end{equation*}
So, combing with (\ref{eq:diff_g-i-k_t_pp}), there is
\begin{equation*}
\begin{aligned}
\bs\sigma^G-\widehat{\bs\sigma^{V}}
=&\bs\sigma^G-\widehat{\bs\sigma^{H}}+\widehat{\bs\sigma^{H}}-\widehat{\bs\sigma^{V}}\\
=&\widehat{\q\otimes\frac{\q}{\rho}}-{\hat\q\otimes\frac{\hat\q}{\hat\rho}} - 1/(8r_s^3)\sum_{\x_{ij}\cap\partial\Omega}\widehat{(\x_{ij}\otimes\f_{ij}c_{ij})}.
\end{aligned}
\end{equation*}
For Piola-Kirchhoff stress, as there is no explicit kinetic term, thus
\begin{equation*}
{\mb P}^G-\widehat{{\mb P}^V}
= - 1/(8r_s^3)\sum_{\X_{ij}\cap\partial\Omega}\widehat{(\X_{ij}\otimes\f_{ij}c_{ij})}.
\end{equation*}
In the above equations, the summation $\sum_{\X_{ij}\cap\partial\Omega}$ is doing on atom $i,j$ such that their $ij$ bond intersects with the boundary of the support of spacial kernel, $c_{ij}$ is the fraction of the $ij$ bond that lies inside of the domain, and the hat ($\ \widehat{}\ $) means convolution with temporal kernel function $\tau$.
\begin{figure}[htbp]
\begin{center}
\includegraphics[scale=0.25]{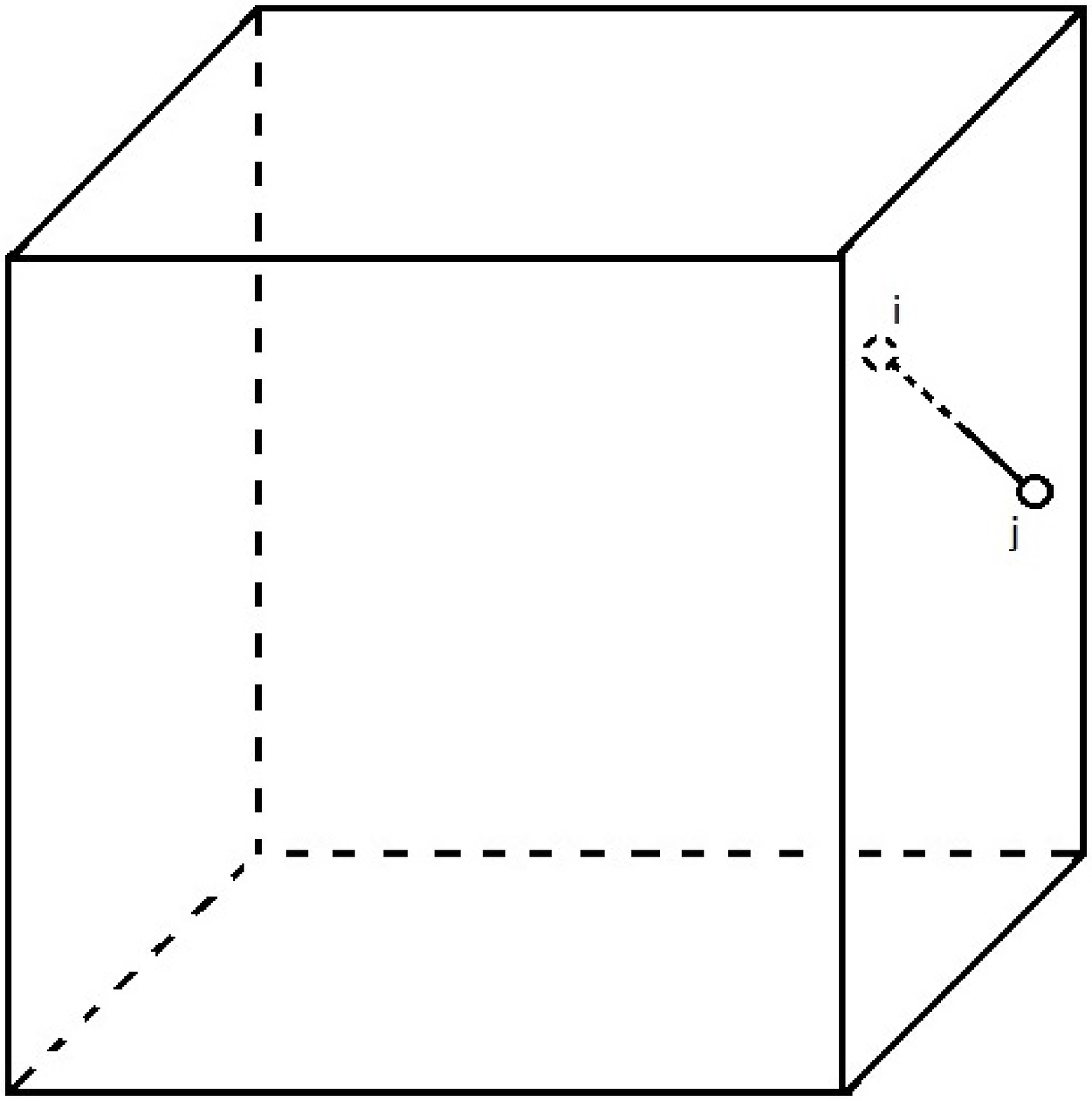}
\caption{The $ij$-bond which intersects with the boundary of the support of spacial kernel.}
\end{center}
\end{figure}

As the summation is doing on the boundary, and the potential function has only finite range of interaction, there holds:
\begin{equation*}
\sum_{\bs\xi_{ij}\cap\partial\Omega}(\bs\xi_{ij}\otimes\f_{ij}c_{ij})=O(r_s^2),\ {\rm where}\ \bs\xi=\x\ {\rm or}\ \X.
\end{equation*}
So
\begin{equation*}
  \begin{aligned}
    \bs\sigma^G-\widehat{\bs\sigma^{V}}=& \widehat{\q\otimes\frac{\q}{\rho}}-{\hat\q\otimes\frac{\hat\q}{\hat\rho}}+O(r_s^{-1}),\\
    {\mb P}^G-\widehat{{\mb P}^V} =& O(r_s^{-1}).
  \end{aligned}
\end{equation*}
We know the part $\widehat{\q\otimes\frac{\q}{\rho}}-{\hat\q\otimes\frac{\hat\q}{\hat\rho}}$ is identical to the difference between g-IK stress and time-averaged Hardy stress. Our analysis above shows that this value vanishes when the spacial kernel radius $r_s$ is large enough.
Therefore we get
\begin{equation}\label{eq:g-i-k_Virial}
\bs\sigma^G\approx\widehat{\bs\sigma^{V}},\ {\mb P}^G\approx\widehat{\mb P^V},\ {\rm when}\ r_s>>1.
\end{equation}
Namely, the virial stress can be derived from g-IK formulism as a special case (kernel has uniform weighting form, and $r_s$ is taken large enough value).

\subsection{PK and Cauchy stress}
In continuum mechanics, there holds the following classical relationship between Cauchy and first Piola-Kirchhoff stress:
\begin{equation}\label{eq:cla_pk_cc}
  {\bs\sigma}=\frac{1}{\det(\mb F)}\mb F\cdot{\mb P}.
\end{equation}
In this subsection, we try to investigate \eqref{eq:cla_pk_cc} in the framework of g-IK formulism.
As is shown above, the g-IK stress becomes virial stress when space-time kernel has uniform weighting form and the spacial radius is large enough. So one way to look at \eqref{eq:cla_pk_cc}
in the framework of g-IK formulism is to study the relationship between $\widehat{\bs\sigma^{V}}$ and $\widehat{\mb P^V}$. Here, we follow the idea used in \cite{zimmerman2010material}. Namely, for each atom $i$, we assume
\begin{equation*}
\mb x_i=\mb F\cdot\mb X_i+\mb z_i,
\end{equation*}
so, the position of each atom $\mb x_i$ is decomposed into a homogeneous deformation $\mb F$ relative
to the material point $\mb X_i$, plus a perturbation due to thermal fluctuations.
Then the virial stress in Eulerian reference can be written as
\begin{equation}\begin{aligned}\label{eq:Virial_L-E}
\bs\sigma^V
    &=\frac{-1}{2|\Omega_0|\det(\mb F)}\sum_{i}\sum_{j\neq i}\mb F\cdot\mb X_{ij}\otimes\mb f_{ij}\\
    &-\frac{1}{2|\Omega|}\sum_{i}\sum_{j\neq i}\mb z_{ij}\otimes\mb f_{ij}
        -\frac{1}{|\Omega|}\sum_im_i\mb v_i\otimes\mb v_i\\
    &=\frac{1}{\det(\mb F)}\mb F\cdot\mb P^V
        -\frac{1}{2|\Omega|}\sum_{i}\sum_{j\neq i}\mb z_{ij}\otimes\mb f_{ij}
        -\frac{1}{|\Omega|}\sum_im_i\mb v_i\otimes\mb v_i,
\end{aligned}\end{equation}
where $\mb z_{ij}={\mb z}_i-{\mb z}_j$, and $|\Omega|=\det(\mb F)|\Omega_0|$.

In zero temperature case, $\mb v_i$ and $\mb z_i$ have zero values,
(\ref{eq:Virial_L-E}) directly gives the relationship between PK and Cauchy stress \eqref{eq:cla_pk_cc}. In finite temperature case, the remaining part including $\mb v_i$ and $\mb z_i$ can be written as
\begin{equation*}\begin{aligned}
&-\frac{1}{2|\Omega|}\sum_{i}\sum_{j\neq i}\mb z_{ij}\otimes\mb f_{ij}
-\frac{1}{|\Omega|}\sum_im_i\mb v_i\otimes\mb v_i\\
&=-\frac{1}{2|\Omega|}\sum_{i}\sum_{j\neq i}(\mb z_{i}-\mb z_j)\otimes\mb f_{ij}
-\frac{1}{|\Omega|}\sum_im_i\mb v_i\otimes\mb v_i\\
&=-\frac{1}{|\Omega|}\sum_{i}\sum_{j\neq i}\mb z_{i}\otimes\mb f_{ij}
-\frac{1}{|\Omega|}\sum_im_i\mb v_i\otimes\mb v_i\\
&=-\frac{1}{|\Omega|}\sum_{i}\mb z_{i}\otimes\mb f_{i}
-\frac{1}{|\Omega|}\sum_im_i\mb v_i\otimes\mb v_i\\
&=-\frac{1}{|\Omega|}\sum_{i}(m_i\mb z_{i}\otimes\frac{d\mb v_{i}}{dt}+m_i\mb v_i\otimes\mb v_i)\\
&=\frac{d}{dt}(-\frac{1}{|\Omega|}\sum_{i}m_i\mb z_i\otimes\mb v_i).
\end{aligned}\end{equation*}
Here we define the notation for atom average and temporal average. For any quantity $\bs\Xi$ related to
atom $i$ and time $t$, we have
\begin{equation*}
\langle\bs\Xi\rangle_{(i)}=\frac{1}{N}\sum_i\bs\Xi_i,\quad
\langle\bs\Xi\rangle_{(t)}=\widehat{\bs\Xi}=\frac{1}{r_t}\int_{t-r_t}^t\bs\Xi(s)ds,
\end{equation*}
thus
\begin{equation*}
\langle{\bs\sigma}^V\rangle_{(t)}-\frac{1}{\det(\mb F)}\mb F\cdot\langle\mb P^V\rangle_{(t)}
    =\frac{N}{|\Omega| r_t}\langle m_i\mb z_i\otimes\mb v_i\rangle_{(i)}\big|_{t}^{t-r_t},
\end{equation*}
then there holds
\begin{equation*}
\big\|\langle{\bs\sigma}^V\rangle_{(t)}-\frac{1}{\det(\mb F)}\mb F\cdot\langle\mb P^V\rangle_{(t)}\big\|_2
    \le\frac{2N}{|\Omega|r_t}\langle\big\|\mb z_i\big\|_2\big\|m_i\mb v_i\big\|_2\rangle_{(i)},
\end{equation*}
then by Cauchy-Schwartz inequality, there holds
\begin{equation*}
\frac{2N}{|\Omega|r_t}\langle\big\|\mb z_i\big\|_2\big\|m_i\mb v_i\big\|_2\rangle_{(i)}
    \le\frac{2N}{|\Omega|r_t}\sqrt{
        \langle\big\|\mb z_i\big\|_2^2\rangle_{(i)}\langle\big\|m_i\mb v_i\big\|_2^2\rangle_{(i)}}
    \le\frac{2N m}{\det(\F)|\Omega_0|r_t}\sqrt{3k_B{\rm T}\Theta},
\end{equation*}
therefore
\begin{equation}\label{eq:Virial_fin-temp}
\big\|\widehat{\bs\sigma^V}-\frac{1}{\det(\mb F)}\mb F\cdot\widehat{\mb P^V}\big\|_2
    \le\frac{2Nm}{\det(\F)|\Omega_0|r_t }\sqrt{3k_B{\rm T}\Theta},
\end{equation}
where $N$ is the number of atoms in $\Omega$, $k_B$ is boltzmann constant, ${\rm T}$ is temperature, $\Theta=\langle\big|\mb z_i\big|_2^2\rangle_{(i)}$ is the mean value of oscillations with respect to all the atoms, and $\dps m=\max_i\{m_i\}$.

From the inequality above, we can see that as $r_t$ increases, the difference between $\widehat{\bs\sigma^V}$ and ${\det(\mb F)}^{-1}\mb F\cdot\widehat{\mb P^V}$ decreases. The temperature is another factor deciding the difference, one extreme case is of the zero temperature, in which case the difference is zero.

From the proof above, we can see the key factor leading to the classical relationship between PK and Cauchy stress is that the oscillation term $\mb z_{i}\otimes\mb f_{i}$ in viscous part of the stress and the kinetic part of stress $m_i\mb v_i\otimes\mb v_i$ cancel with each other in the sense of a temporal average. After cancellation, the remaining viscous part in Cauchy stress can now be directly related to Piola-Kirchhoff stress. In the numerical experiment part, we will further investigate the behaviors of $\mb z_{i}\otimes\mb f_{i}$ and $m_i\mb v_i\otimes\mb v_i$ under the effect of temperature and temporal average.

\subsection{Generalization to general kernel}
The relationship ${\bs\sigma}={\det(\mb F)}^{-1}\mb F\cdot{\mb P}$ could actually be generalized to the case of general shape of kernel function. However, the generalization would make it difficult for us to directly estimate the difference in an explicit formula, as we need to consider the properties of the kernel function as well, which will make the work too much to all fit in this paper. Here, we briefly introduce the generalization to general shape of spacial kernel function in zero temperature.
To proceed, we first need several natural assumptions:
\begin{enumerate}
\item
Finite range interatomic interaction. Namely,
\begin{equation*}
\f_{ij}=0,\ {\rm when}\ \|\x_{ij}\|_2\ge k_0,\ {\rm for\ some\ }k_0.
\end{equation*}
\item
One order regularity of original spacial kernel function (notice that $\varphi(\x)=\frac{1}{r_s^3}\phi(\frac{\x}{r_s})$, and $\varphi$ is the spacial kernel). Namely,
\begin{equation*}
\sup_{\x}\|\nabla\phi(\x)\|_2\le k_1,\ {\rm for\ some\ }k_1.
\end{equation*}
\item
Translational invariance. So there hold,
\begin{equation*}
\sum_{j\neq i}\mb X_{ij}\otimes\mb f_{ij}={\mb M},\ \forall\ i,\ {\rm for\ some\ }\mb M.
\end{equation*}
\end{enumerate}

Now we consider a lattice under a uniform deformation $\mb F$, namely $\x_i=\mb F\cdot\X_i$. When the kernel is space-time separable, the g-IK stress is equal to the Hardy stress no matter what kind of temporal average is taken. So we will directly compare the two versions (Cauchy and PK) of Hardy stress rather than g-IK stress. For the Cauchy stress in zero temperature, there is
\begin{equation*}\begin{aligned}
\bs\sigma^H
    &=\frac{-1}{2}\sum_{i}\sum_{j\neq i}\mb x_{ij}
        \otimes\mb f_{ij}\int_0^1\varphi\left(\mb x_i+\lambda\mb x_{ji}\right)d\lambda.
\end{aligned}\end{equation*}
Here we assume the sample point is at 0. By usage of assumption 1 and the fact that the support of $\varphi$ has radius of $r_s$, we know the summation $\sum_{i}\sum_{j\neq i}$ has the following transform
\begin{equation*}
\sum_{i}\sum_{j\neq i}=\sum_{i\in \mathcal B}\sum_{j\in \mathcal D_i},
\end{equation*}
where
\begin{equation*}
\begin{aligned}
\mathcal B:=\{i:\ \x_i\in B(0,r_s+k_0)\},\\
\mathcal D_i:=\{j:\ i\neq j,\ \|\x_{ij}\|\le k_0\}.
\end{aligned}
\end{equation*}
So we have
\begin{equation*}\begin{aligned}
\bs\sigma^H
    &=\frac{-1}{2}\sum_{i\in \mathcal B}\sum_{j\in \mathcal D_i}\mb F\cdot\mb X_{ij}
        \otimes\mb f_{ij}\int_0^1\varphi\left(\mb F\cdot(\mb X_i+\lambda\mb X_{ji})\right)d\lambda\\
    &=\frac{-1}{2}\sum_{i\in \mathcal B}\sum_{j\in \mathcal D_i}\mb F\cdot\mb X_{ij}
        \otimes\mb f_{ij}\int_0^1\frac{1}{r_s^3}\phi\left(\frac{\mb F\cdot\mb X_i}{r_s}+\frac{\lambda\mb F\cdot\mb X_{ji}}{r_s}\right)d\lambda.
\end{aligned}\end{equation*}
Now, from assumption 1 and 2, we have
\begin{equation*}
\phi\left(\frac{\mb F\cdot\mb X_i}{r_s}+\frac{\lambda\mb F\cdot\mb X_{ji}}{r_s}\right)=\phi(\frac{\mb F\cdot\mb X_i}{r_s})+O(r_s^{-1}).
\end{equation*}
So there hold
\begin{equation*}\begin{aligned}
\bs\sigma^H
    &=\frac{-1}{2}\sum_{i\in \mathcal B}\sum_{j\in \mathcal D_i}\mb F\cdot\mb X_{ij}
        \otimes\mb f_{ij}\left[\frac{1}{r_s^3}\phi(\frac{\mb F\cdot\mb X_i}{r_s})+O(r_s^{-4})\right].
\end{aligned}\end{equation*}
As $\#\mathcal B=O(r_s^3)$ and $\sum_{j\in \mathcal D_i}\mb F\cdot\mb X_{ij}\otimes\mb f_{ij}=O(1)$, so
\begin{equation*}
\sum_{i\in \mathcal B}\sum_{j\in \mathcal D_i}\mb F\cdot\mb X_{ij}\otimes\mb f_{ij}=O(r_s^3),
\end{equation*}
thus we have
\begin{equation*}\begin{aligned}
\bs\sigma^H
    &=\frac{-1}{2}\sum_{i}\sum_{j\neq i}\mb F\cdot\mb X_{ij}
        \otimes\mb f_{ij}\frac{1}{r_s^3}\phi(\frac{\mb F\cdot\mb X_i}{r_s})+O(r_s^{-1}).
\end{aligned}\end{equation*}
Similarly, there hold the following equation for the PK stress
\begin{equation*}
\mb P^H=\frac{-1}{2}\sum_{i}\sum_{j\neq i}\mb X_{ij}
        \otimes\mb f_{ij}\frac{1}{r_s^3}\phi(\frac{\mb X_i}{r_s})+O(r_s^{-1}).
\end{equation*}
Now by assumption 3, we have
\begin{equation*}
\sum_{j\neq i}\mb X_{ij}\otimes\mb f_{ij}={\mb M},\ \forall\ i.
\end{equation*}
Then the Cauchy and PK stress become
\begin{equation*}
\begin{aligned}
\bs\sigma^H=&\mb F\cdot\mb M\sum_i\frac{1}{r_s^3}\phi(\frac{\mb F\cdot\mb X_i}{r_s})+O(r_s^{-1}),\\
\mb P^H=&\mb M\sum_i\frac{1}{r_s^3}\phi(\frac{\mb X_i}{r_s})+O(r_s^{-1}).
\end{aligned}
\end{equation*}
Now their relationship becomes clear. This is because when $r_s$ is large enough, the summation above could also be interpreted as integration:
\begin{equation*}
\sum_i\frac{1}{r_s^3}\phi(\frac{\mb F\cdot\mb X_i}{r_s})\approx k\int_{\real^3}\phi(\mb F\cdot\x)dx,\quad
\sum_i\frac{1}{r_s^3}\phi(\frac{\mb X_i}{r_s})\approx k\int_{\real^3}\phi(\x)dx.
\end{equation*}
So, there hold
\begin{equation*}
\bs\sigma^H=\frac{1}{\det(\mb F)}\mb F\cdot\mb P^H,\ {\rm as}\ r_s>>1.
\end{equation*}

\section{Numerical experiments}\label{sec:num_exp}
In this section, we describe several numerical experiments of molecular dynamics,
conducted to further understand our analysis results.
We consider a molecular dynamics model of FCC aluminum (Al).
Atoms are assumed to be interacting through the embedded-atom model (EAM) \cite{daw1984embedded},
in which the potential energy is given by,
\begin{equation}
  V=\frac{1}{2}\sum_{i,j}\phi(r_{ij})+\sum_iU(\rho_i), \quad \rho_i=\sum_{j\ne i}\rho(r_{ij}).
\end{equation}
Here $\phi$ is a pairwise potential, $U$ is the glue function and $\rho$ is the electron density function of the
$i$-th atom. Parameters in the expression can be found in \cite{ercolessi1994interatomic}.
For the EAM model, the force decomposition \(\mb f_{ij}\) is given by,
\begin{equation}\label{eq:eam-fij}
\mb f_{ij}= -\Bigl[ \phi'(r_{ij}) + U'(\rho_i) + U'(\rho_j)\Bigr] \frac{\mb r_{ij}}{r_{ij}}.
\end{equation}

In the MD simulations, we use the standard neighbor list method \cite{AlTi89,FrSm02} in the force calculation.
We use the NVE ensemble, in which the standard Verlet's time integrator is used.
The lattice constant for such a system is \(a_0=4.032\)\AA\ at zero temperature and \(a_0=4.051\)\AA\ at $300$K.
The energy unit is in \(eV\). The time scale is \(0.052880\) pico-second and the unit for the stress is \(160.2176\) GPa. All the results will be presented in these unit. The step size for the time integration is \(\Delta t=0.1\) for experiment 1 and \(\Delta t=0.01\) for experiment 2.

\subsection{Experiment 1}
Our first experiment is on the difference between g-IK stress and time-averaged Hardy stress when the kernel function is space-time separable. As we show in the previous section, there hold (\ref{eq:diff_g-i-k_t_pp}), namely
\begin{equation*}
\widehat{\bs\sigma^{H}}(\x,t)-\bs\sigma^{G}(\x,t)=\widehat{\q\otimes\frac{\q}{\rho}}-{\hat\q\otimes\frac{\hat\q}{\hat\rho}}.
\end{equation*}
Thus the estimation of the difference between the two versions of stress can be translated to the estimation of the difference between $\widehat{\q\otimes\frac{\q}{\rho}}$ and ${\hat\q\otimes\frac{\hat\q}{\hat\rho}}$. Here we will call $\widehat{\q\otimes\frac{\q}{\rho}}$ as the time-averaged part, and ${\hat\q\otimes\frac{\hat\q}{\hat\rho}}$ as the g-IK part.

\begin{figure}[htbp]
\begin{center}
\includegraphics[scale=0.35]{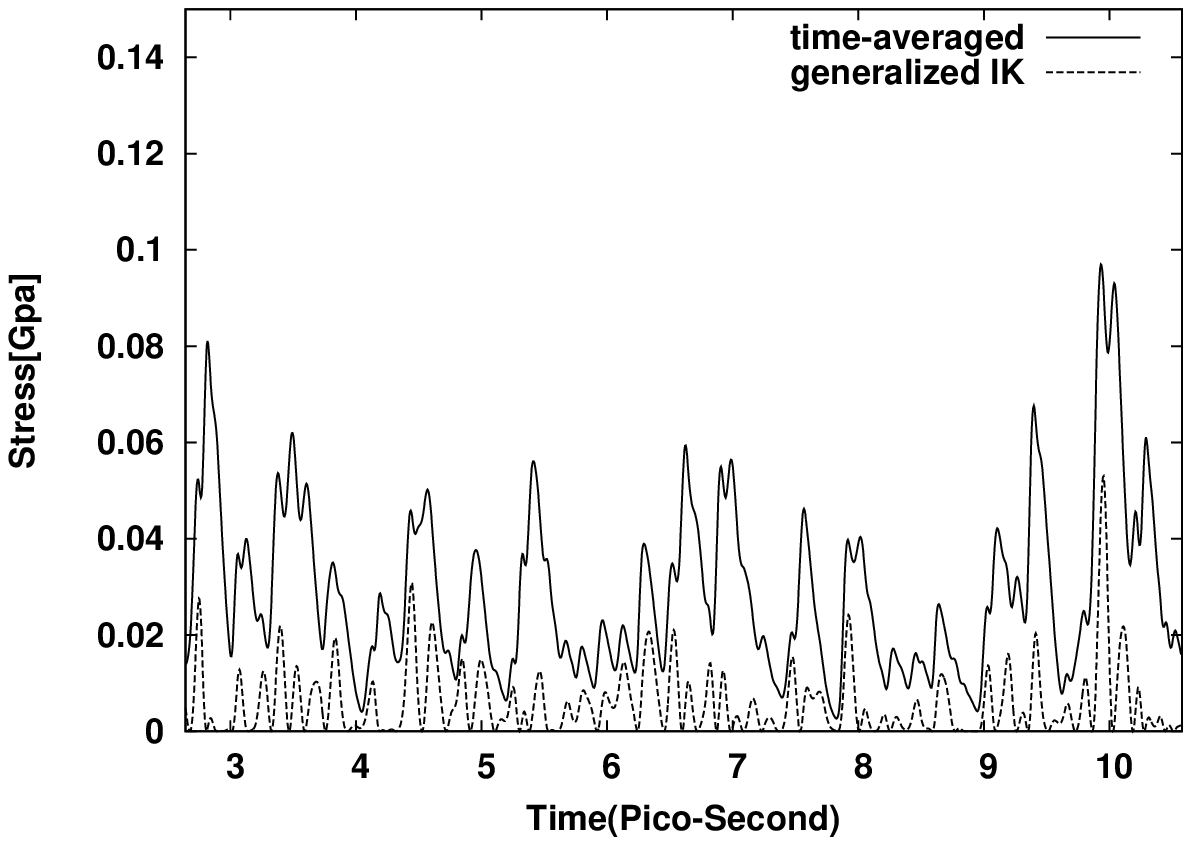}
\includegraphics[scale=0.35]{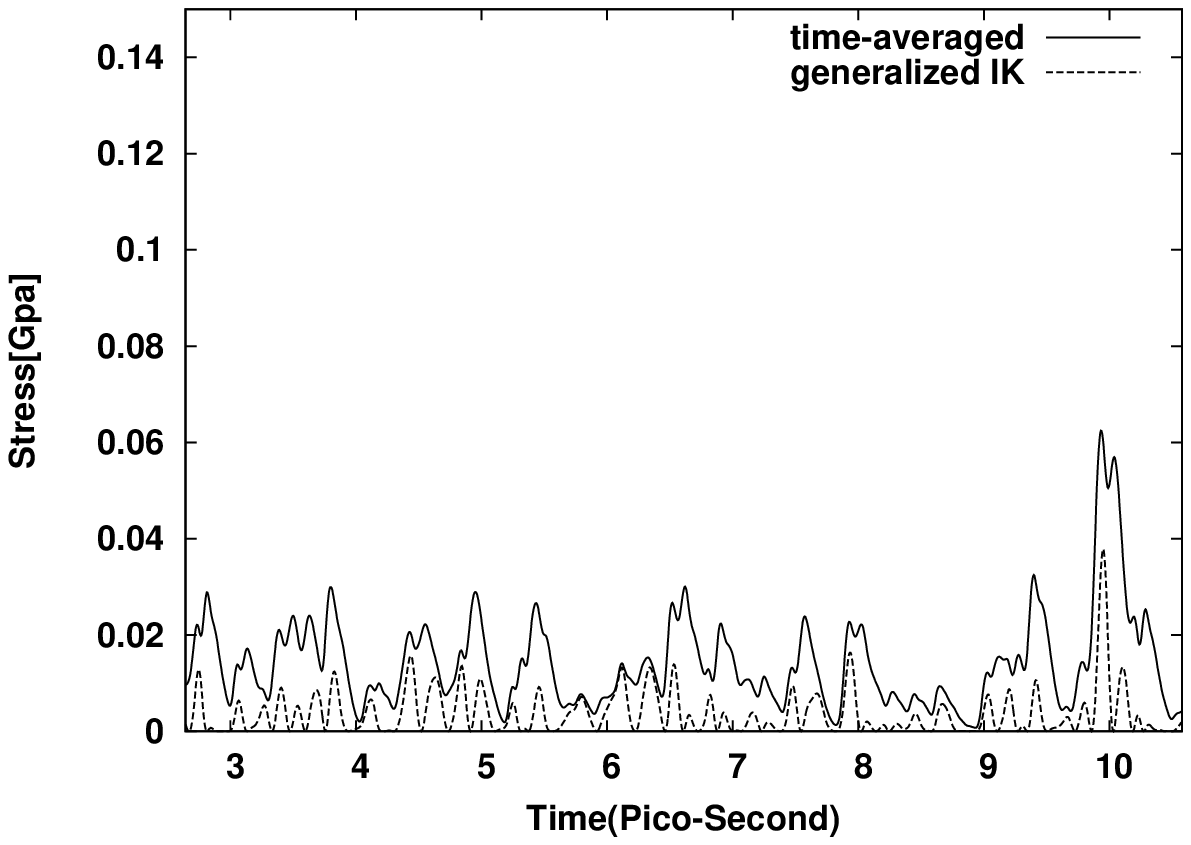}
\includegraphics[scale=0.35]{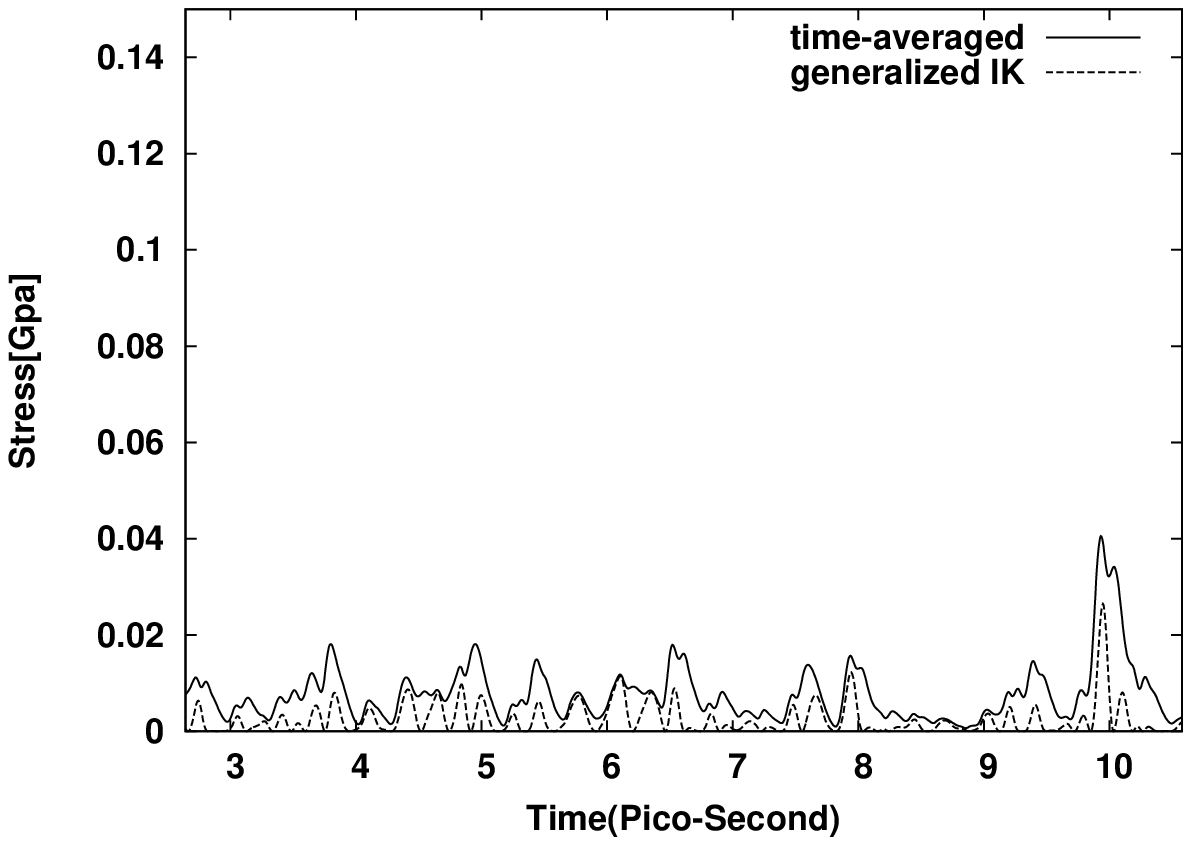}
\includegraphics[scale=0.35]{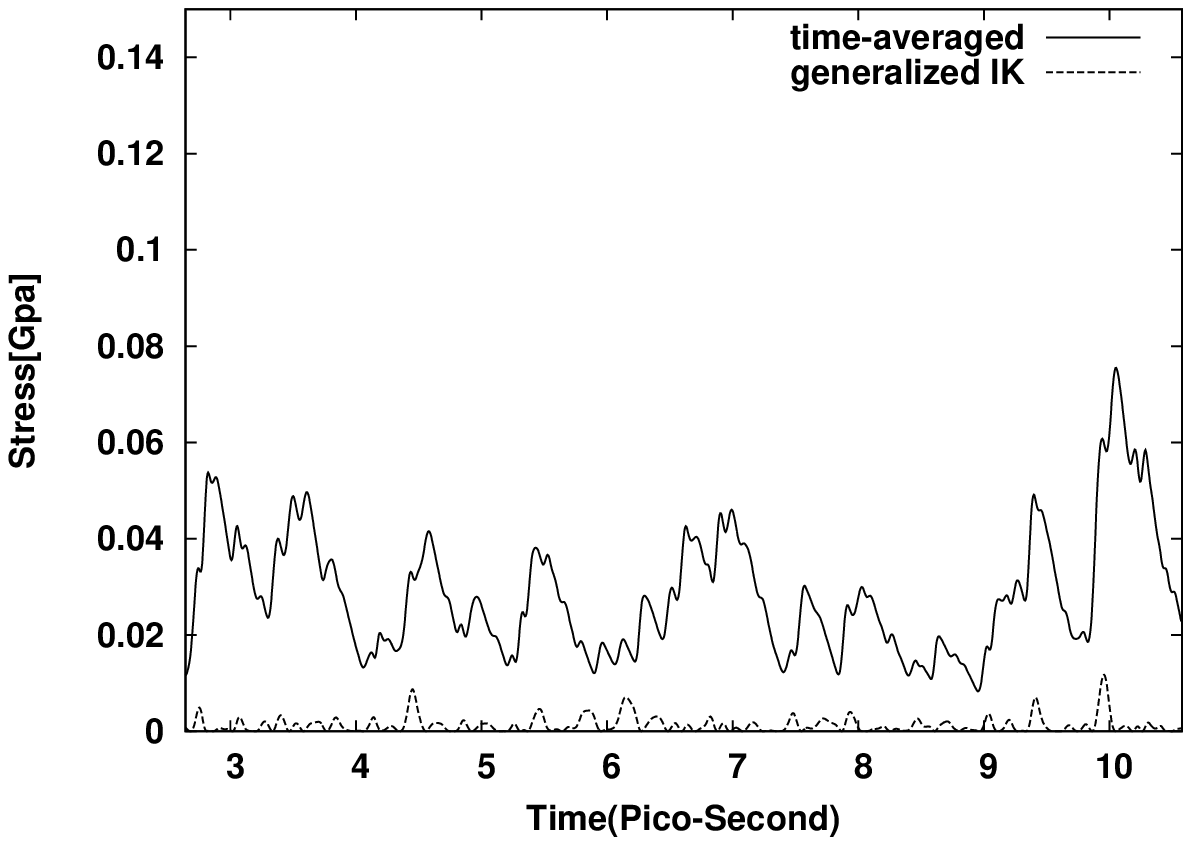}
\includegraphics[scale=0.35]{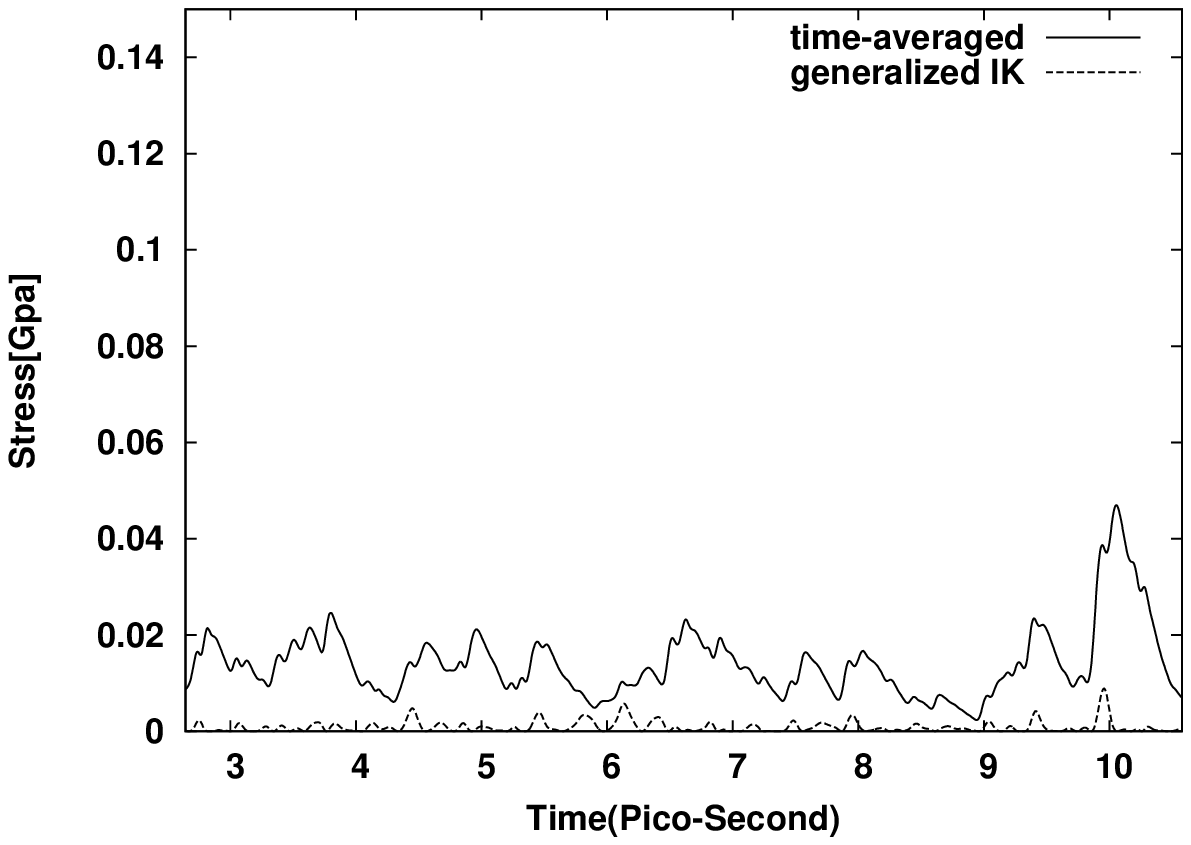}
\includegraphics[scale=0.35]{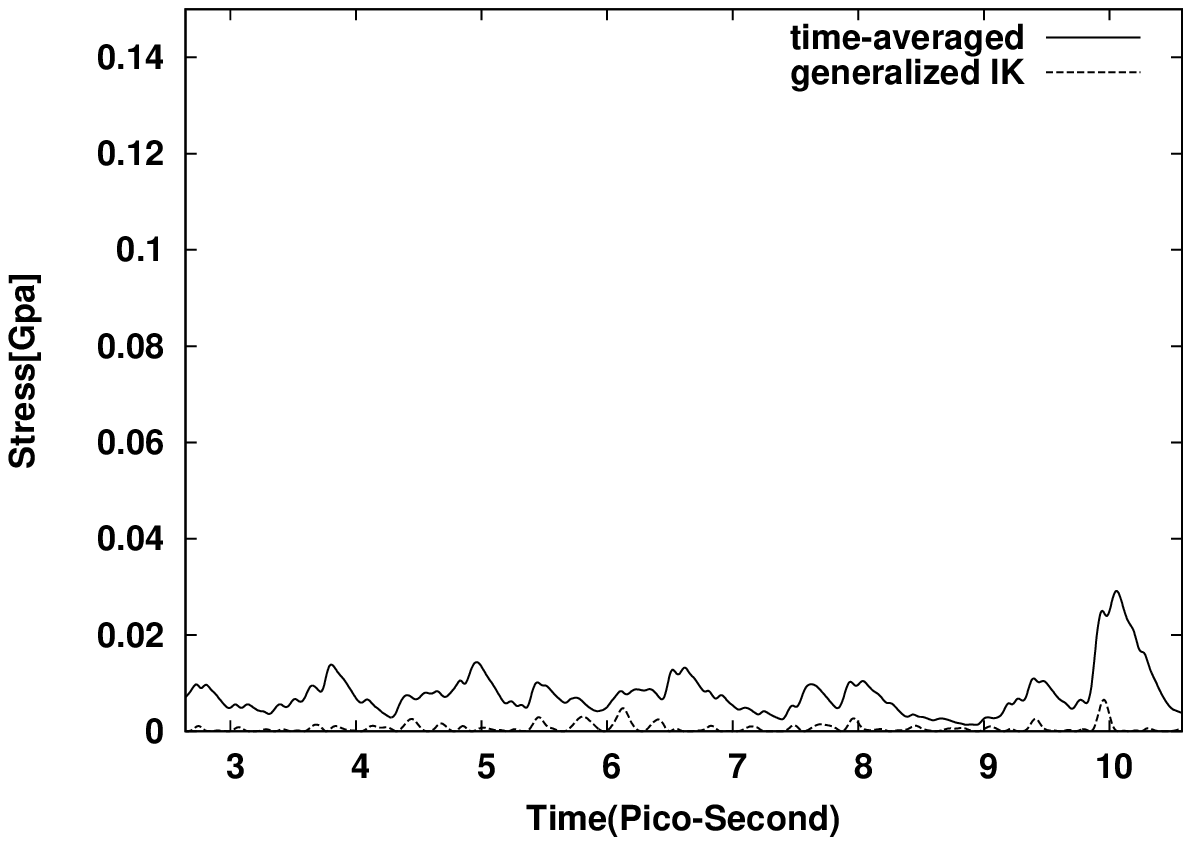}
\includegraphics[scale=0.35]{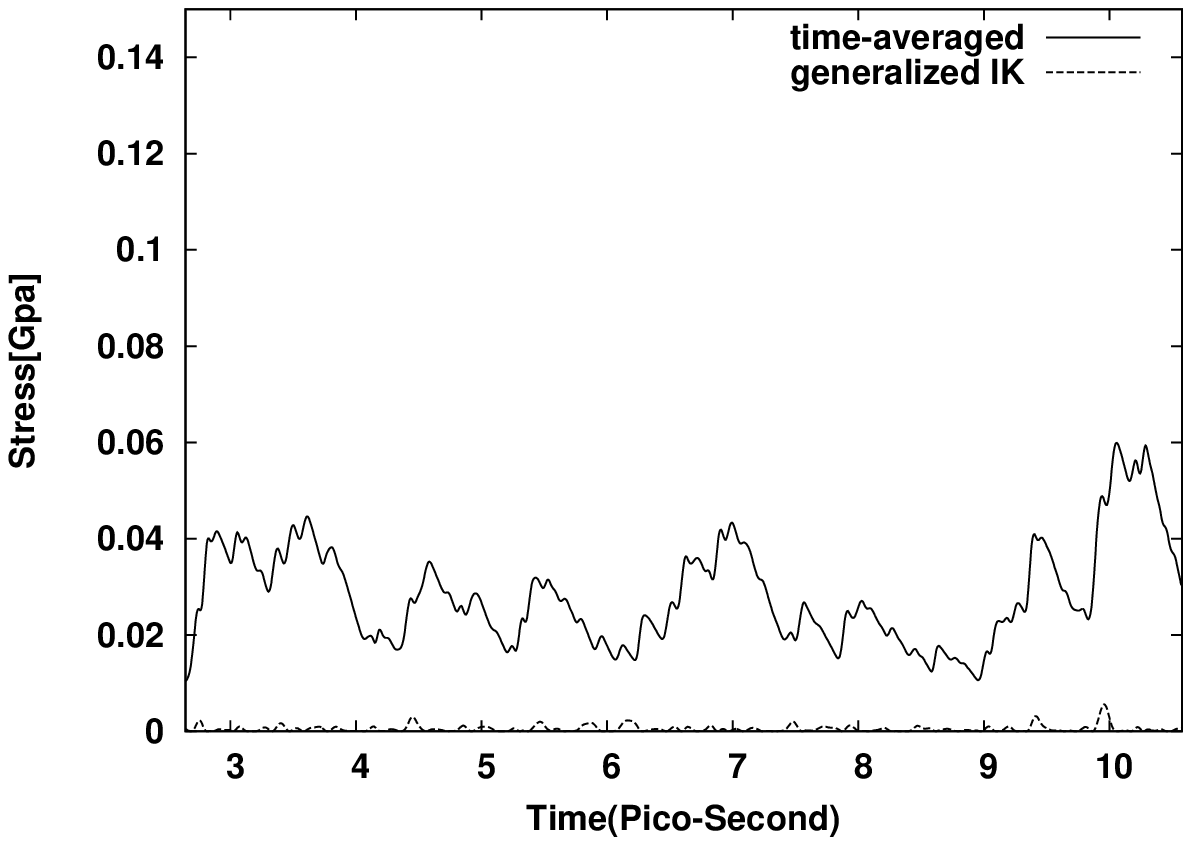}
\includegraphics[scale=0.35]{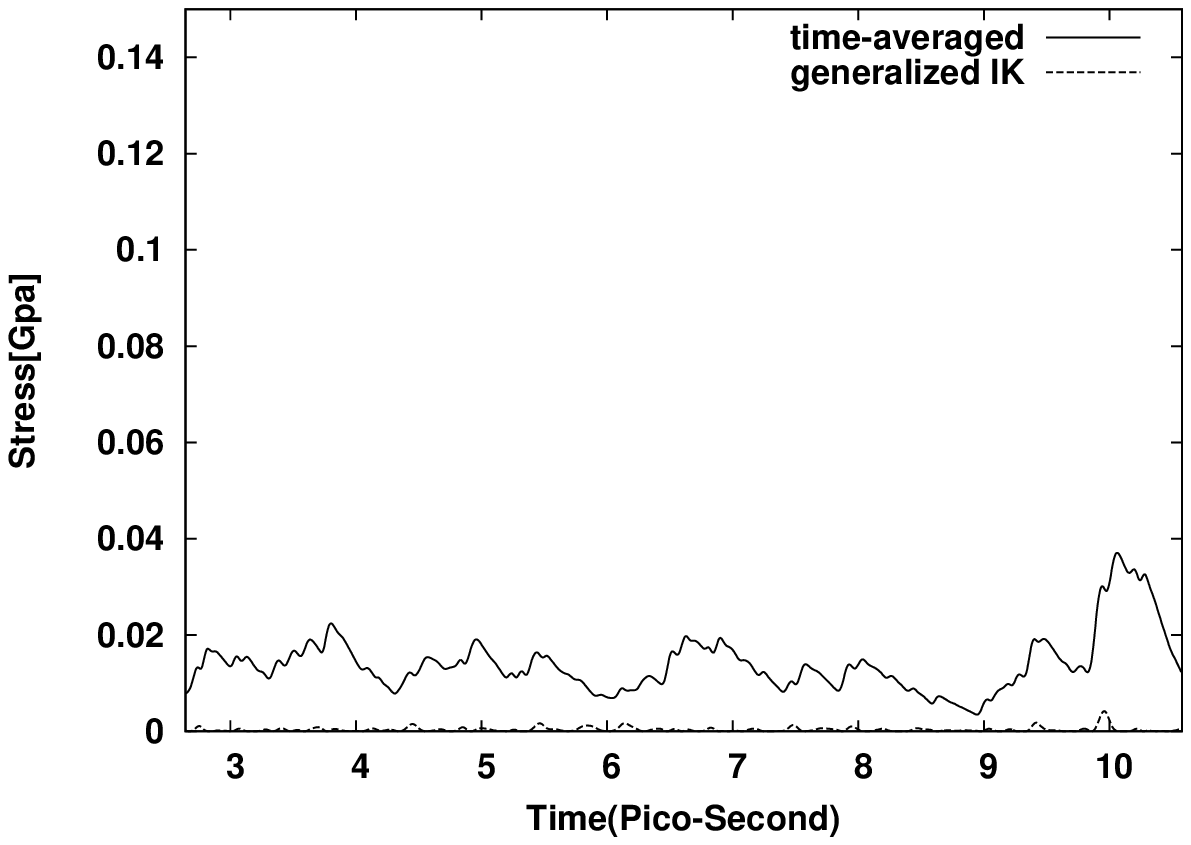}
\includegraphics[scale=0.35]{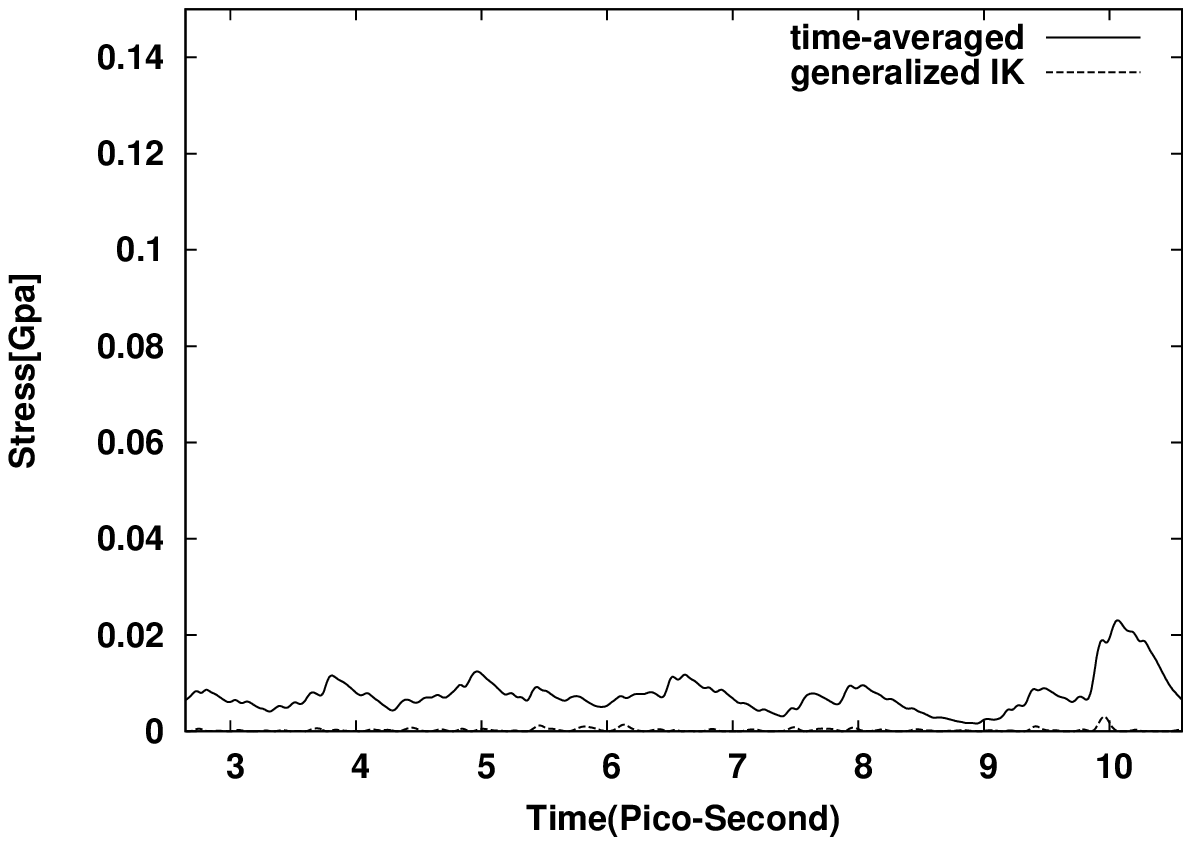}
\end{center}
\caption{Variation of $(\widehat{\q\otimes\frac{\q}{\rho}})_{11}$ and $({\hat\q\otimes\frac{\hat\q}{\hat\rho}})_{11}$. Kernel $\Psi^\textrm{I}$ is used. From left to right, $r_s=$4.032\AA, 5.04\AA, 6.048\AA. From top to down, $r_t$=2.644, 5.288, 7.932 Pico-Seconds.}
\label{fig:ta_gik}
\end{figure}

We set up the system under the temperature of $300$K and track the variations of the two parts as MD system evolves. Figure \ref{fig:ta_gik} shows the variations of both parts under different spacial and temporal radius. We can see both parts go down when spacial radius becomes larger. When temporal radius becomes larger, we can see the time-averaged part becomes more stable, which is a reasonable effect of temporal average, but its mean value nearly keeps unchanged. For the g-IK part, we see a significant decrease in whole. So, as a consequence, their difference becomes larger when temporal radius increases.

The numerical results are consistent with our analysis. As the MD system is in thermodynamics equilibrium. When the spacial radius becomes larger. The physical quantities $\q,\v$ become more closer to the system's global value (which is constant zero by our initiation) and more stable to time. When temporal radius becomes larger, $\hat\q, \hat\v$ become more closer to the global value as an effect of ensemble or temporal average, but $\widehat{\q\otimes\frac{\q}{\rho}}$ does not decrease for the components $\{(\q\otimes\q)_{ii},\ i=1,2,3\}$ are always positive values. This will make the oscillations unable to cancel, but accumulate over temporal average.

\begin{figure}[htbp]
\begin{center}
\includegraphics[scale=0.7]{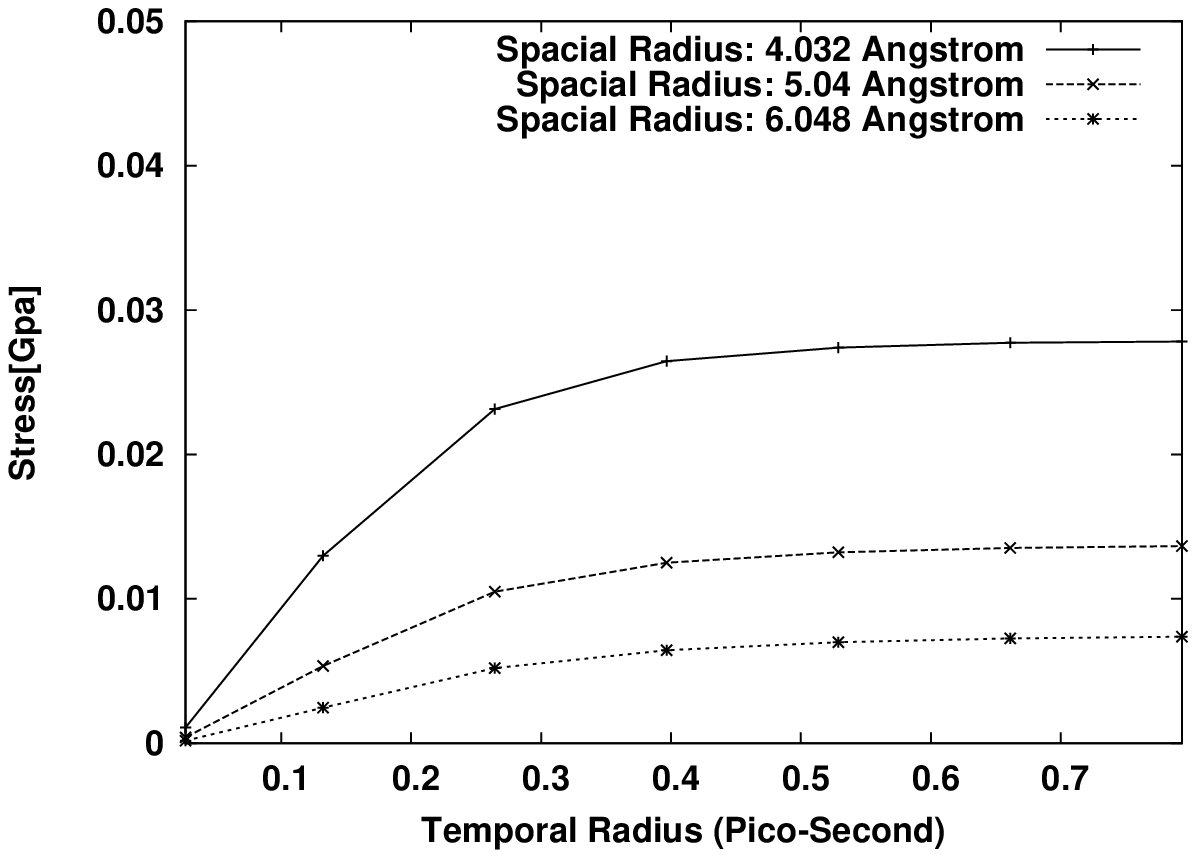}
\caption{The mean value of $|\widehat{\bs\sigma_{11}^{H}}-\bs\sigma^{G}_{11}|$,
estimated under different spacial and temporal radius.
Kernel $\Psi^\textrm{I}$ is used.}
\label{fig:ta_gik_abs}
\end{center}
\end{figure}

To get a more clear observation on the difference, we draw Figure \ref{fig:ta_gik_abs}. The differences are first valued at each time-step, then we take the mean absolute value of them. This process is repeated under different spacial and temporal radius.

From Figure \ref{fig:ta_gik_abs}, we see the difference decreases to zero when the temporal radius decreases to zero. When temporal radius increases, the difference first increases significantly, then becomes stable at some value. When spacial radius becomes larger, the difference decreases in whole. The experiment results are consistent with our analysis in B of section {\ref{sec:g-IK-t-tpp}}.

\subsection{Experiment 2}
The second experiment is on the relationship between Cauchy and PK stress. From (\ref{eq:Virial_L-E}), we know
\begin{equation*}
\bs\sigma^V
    =\frac{1}{\det(\mb F)}\mb F\cdot\mb P^V
        -\frac{1}{|\Omega|}\sum_i[\mb z_{i}\otimes\mb f_{i}+m_i\mb v_i\otimes\mb v_i],
\end{equation*}
so $\frac{1}{|\Omega|}\sum_i\mb z_{i}\otimes\mb f_{i}$ and $\frac{1}{|\Omega|}\sum_i m_i\mb v_i\otimes\mb v_i$ play important roles on the difference between revised PK stress $({\det(\mb F)}^{-1}\mb F\cdot\mb P)$ and Cauchy stress $(\bs\sigma^V)$, which we suppose to be zero in continuum theory. In the following content,
we will study the behaviors of $\frac{1}{|\Omega|}\sum_i\mb z_{i}\otimes\mb f_{i}$ and $\frac{1}{|\Omega|}\sum_i m_i\mb v_i\otimes\mb v_i$ under different temporal average radius and system temperatures.
We will call $\frac{1}{|\Omega|}\sum_i\mb z_{i}\otimes\mb f_{i}$ as viscous term, and $\frac{1}{|\Omega|}\sum_i m_i\mb v_i\otimes\mb v_i$ as kinetic term.

\begin{figure}[htbp]
\begin{center}
\includegraphics[scale=0.35]{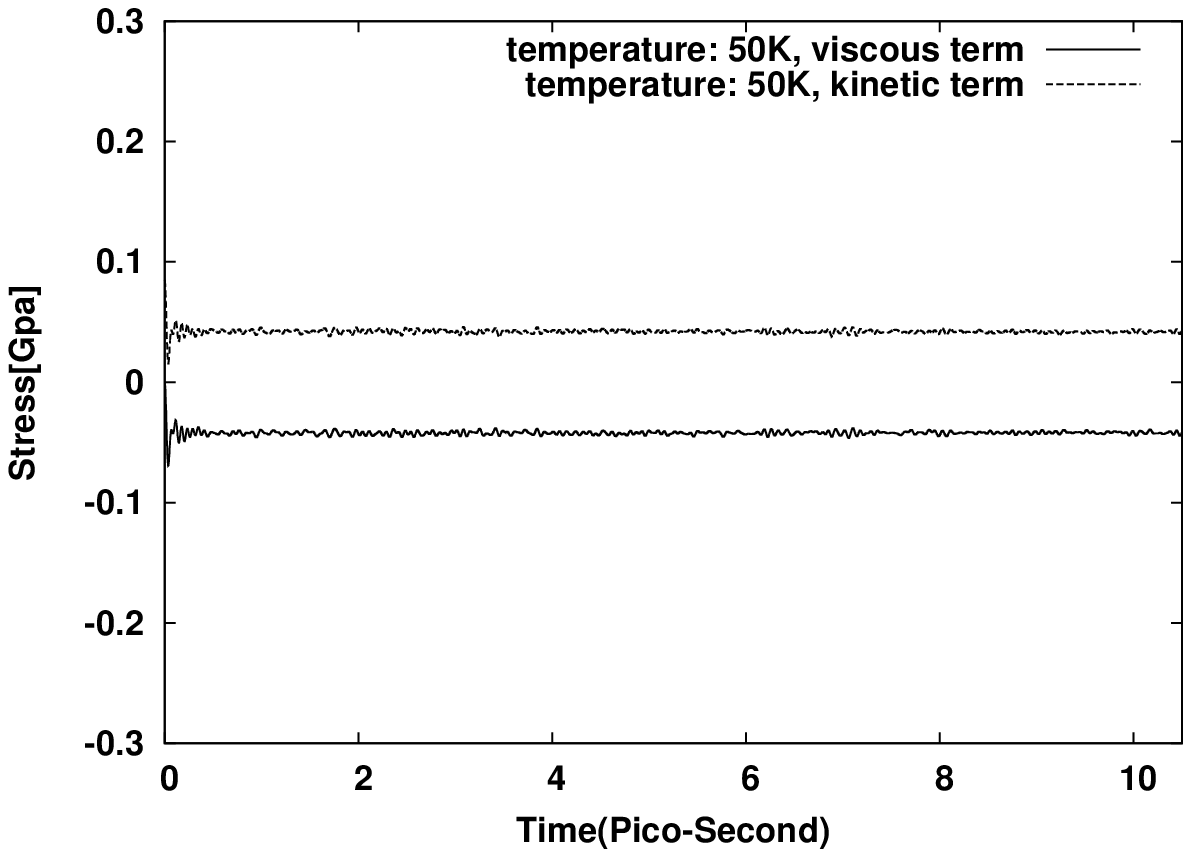}
\includegraphics[scale=0.35]{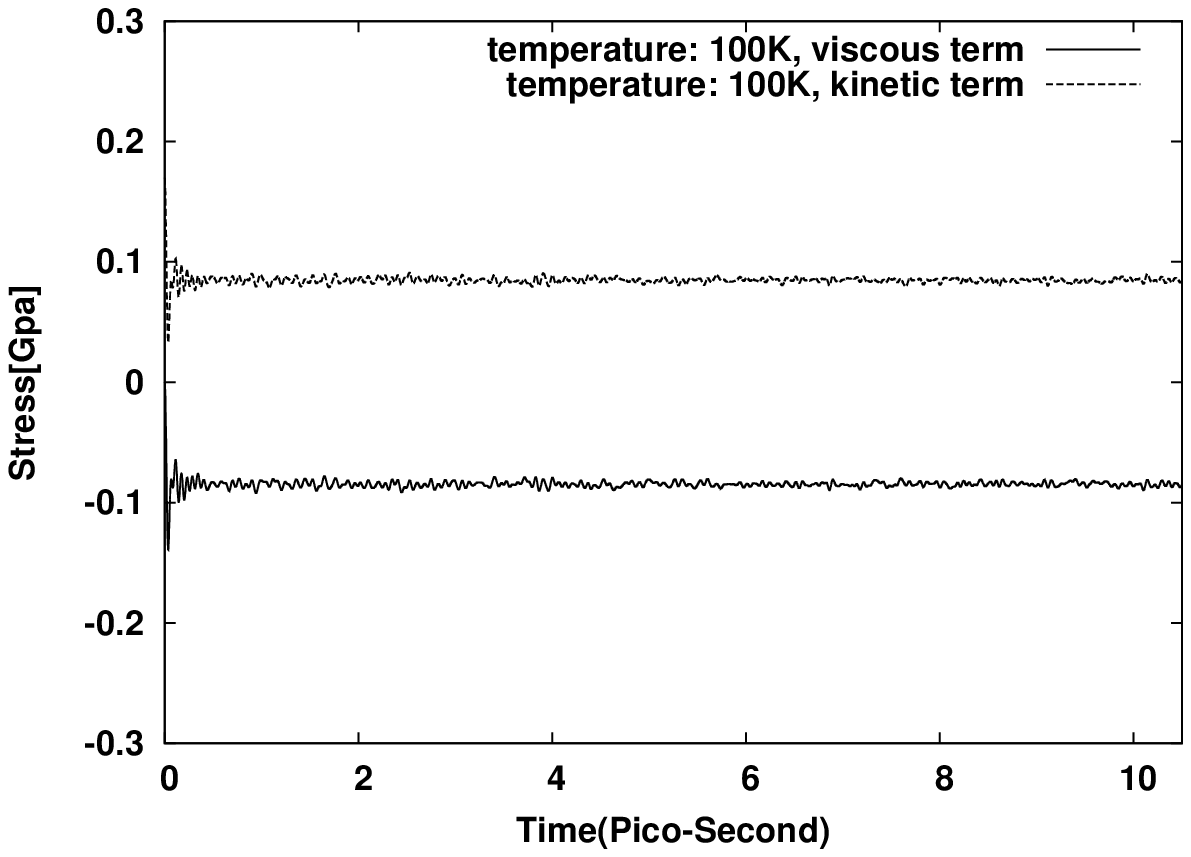}
\includegraphics[scale=0.35]{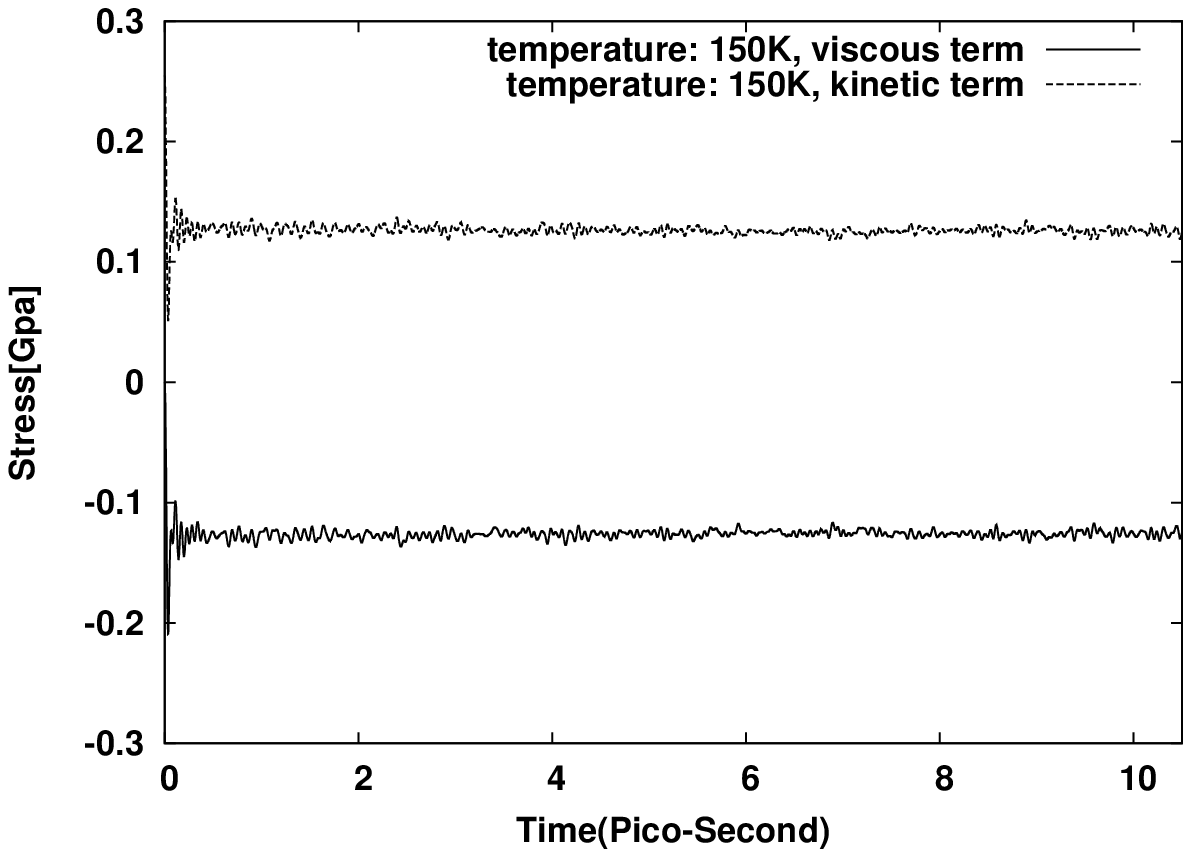}
\end{center}
\caption{Variations of viscous term $(\frac{1}{|\Omega|}\sum_i\mb z_{i}\otimes\mb f_{i})_{11}$ and kinetic term $(\frac{1}{|\Omega|}\sum_i m_i\mb v_i\otimes\mb v_i)_{11}$ at different temperatures. From left to right, the temperatures are 50K, 100K, 150K.}
\label{fig:v_k_sep}
\end{figure}

We first set up the MD system at zero temperature (\(a_0=4.032\)\AA), then we heat the system up to 50K, 100K, 150K and calculate the viscous term and kinetic term respectively at these temperatures. Figure \ref{fig:v_k_sep} shows the variations of the two terms at different temperatures. We can see as the temperature increases, the absolute values of both the two terms increase. This is because the molecules oscillating more fiercely in higher temperature. Another thing worth noticing is that the two terms have nearly the same absolute values despite the different temperatures. As their signs are different, the canceling of the two terms will make their summation varying around zero value.

\begin{figure}[htbp]
\begin{center}
\includegraphics[scale=0.35]{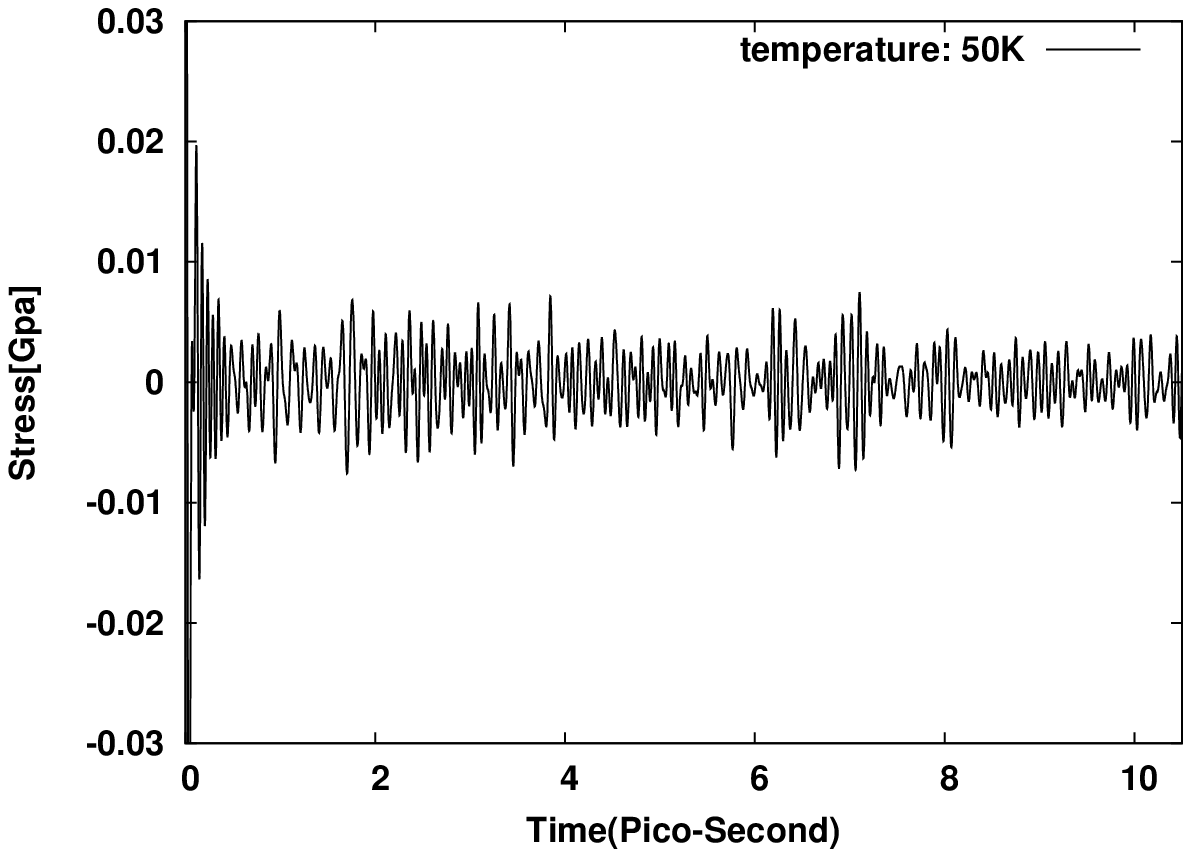}
\includegraphics[scale=0.35]{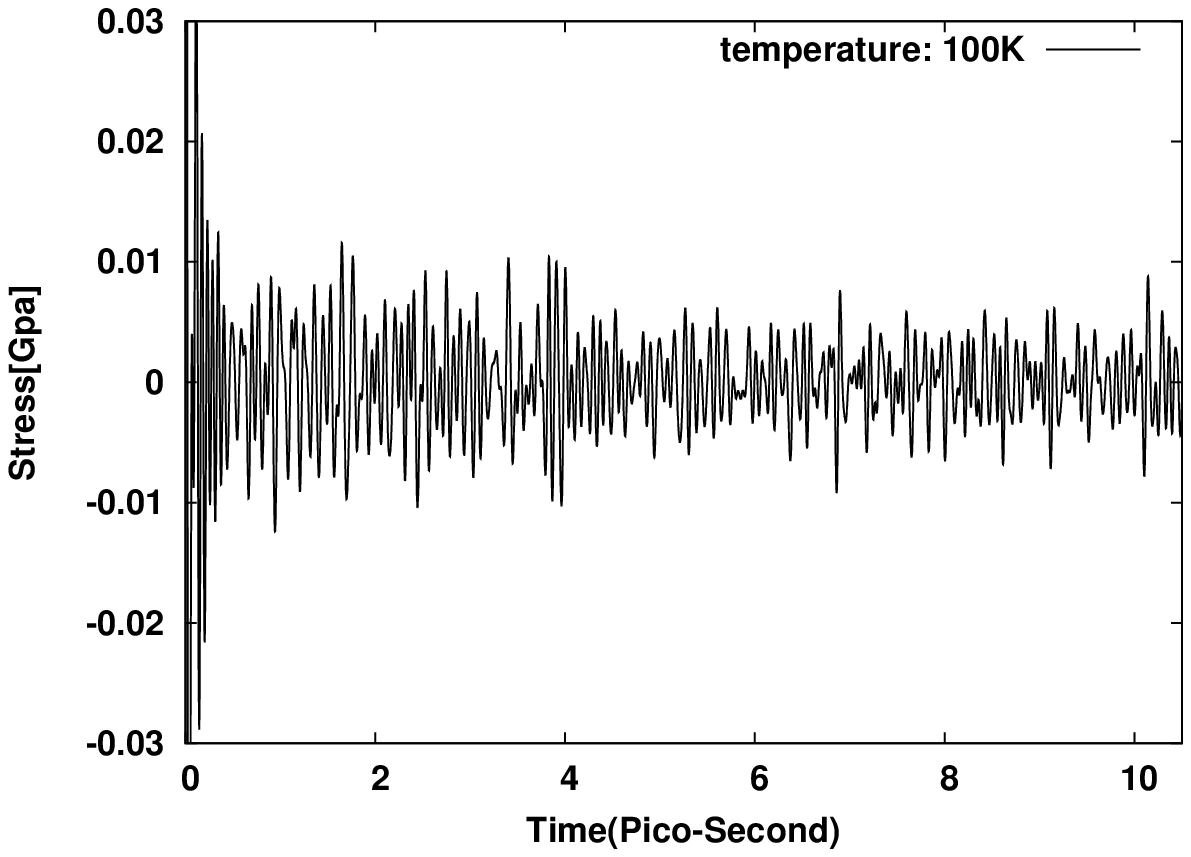}
\includegraphics[scale=0.35]{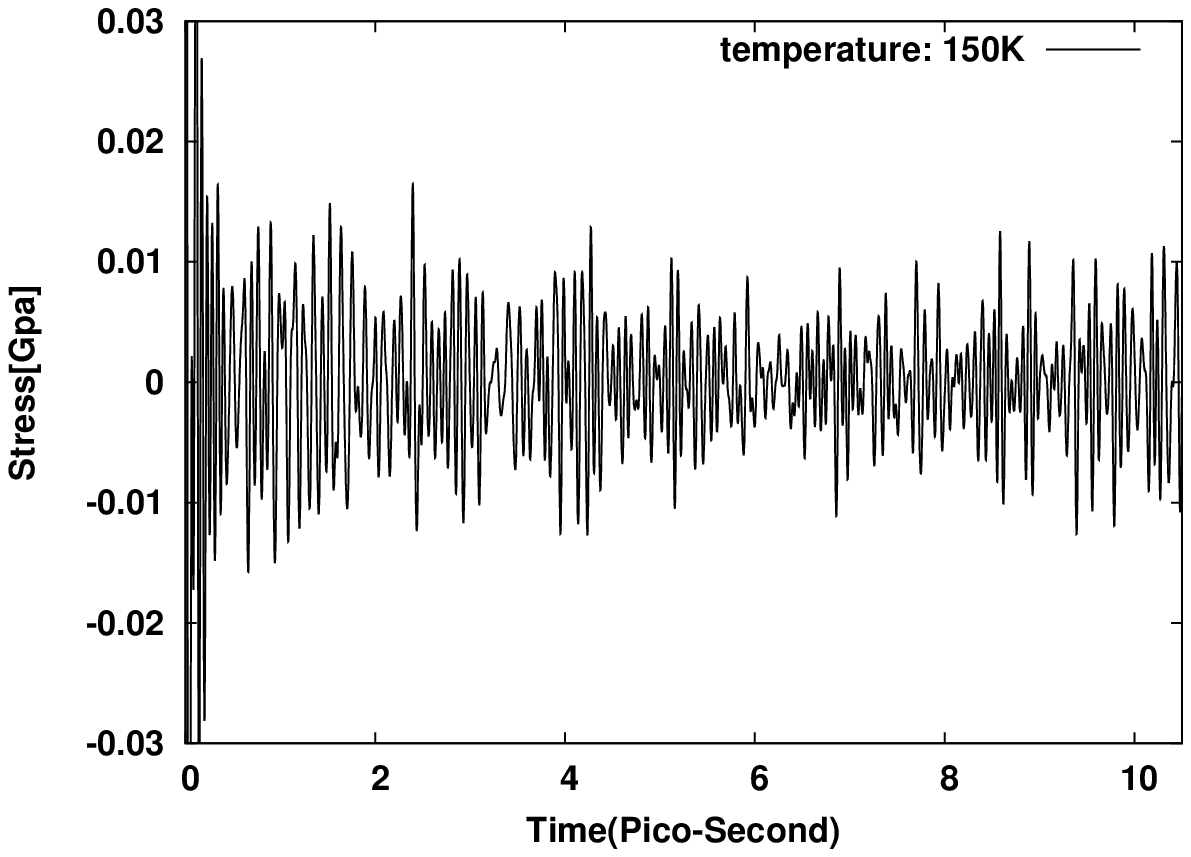}
\includegraphics[scale=0.35]{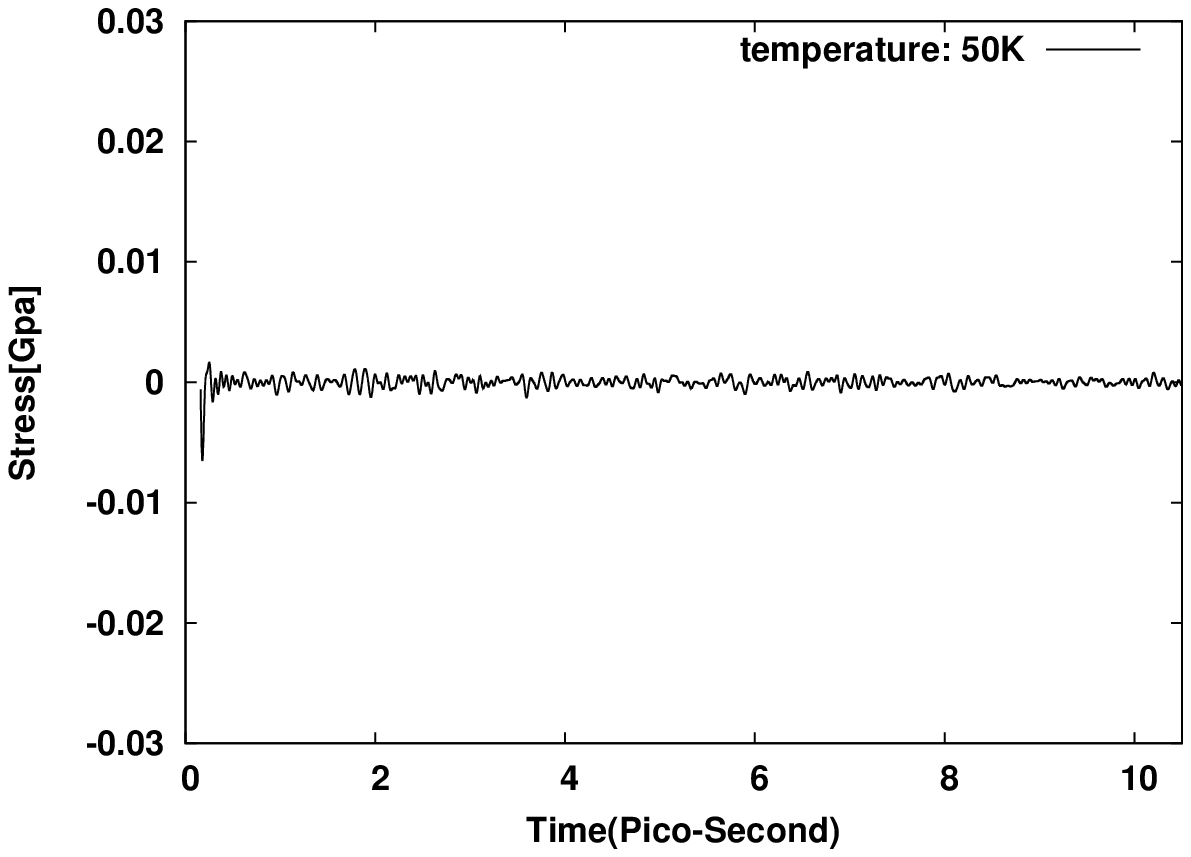}
\includegraphics[scale=0.35]{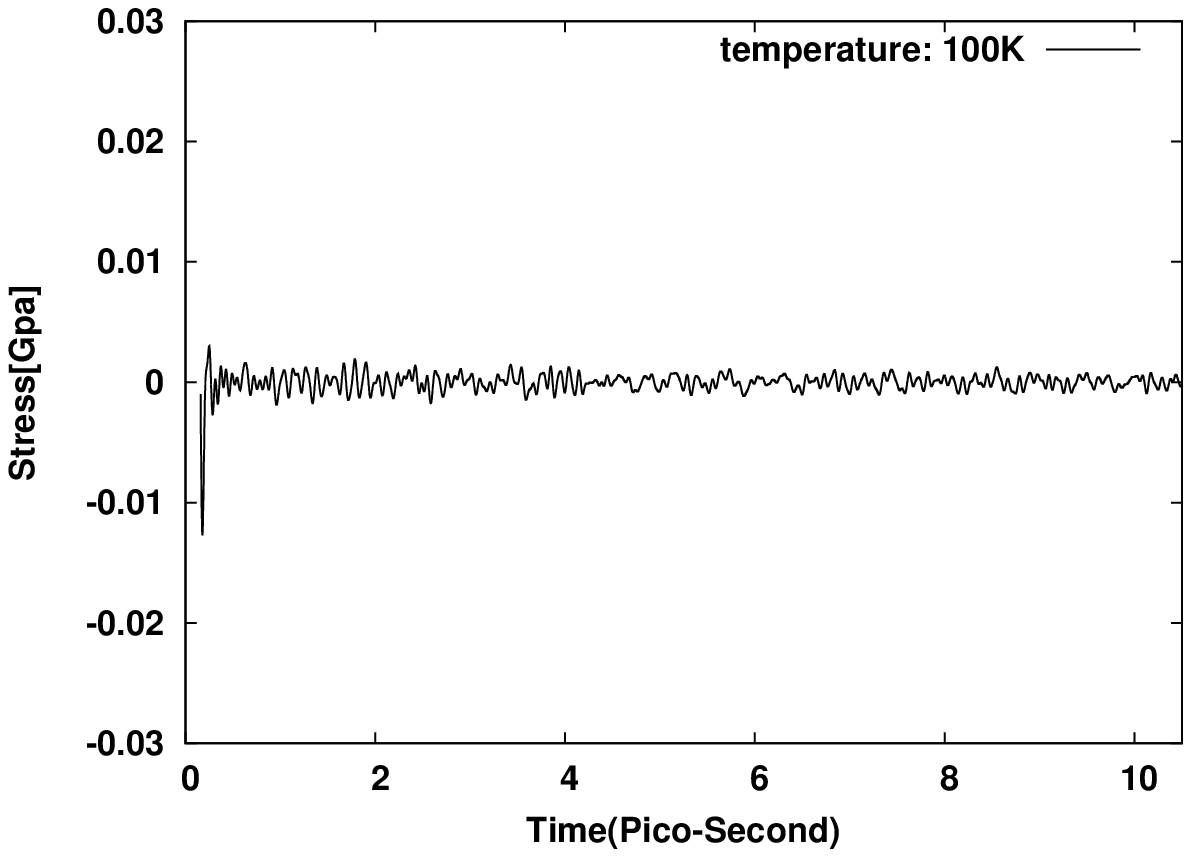}
\includegraphics[scale=0.35]{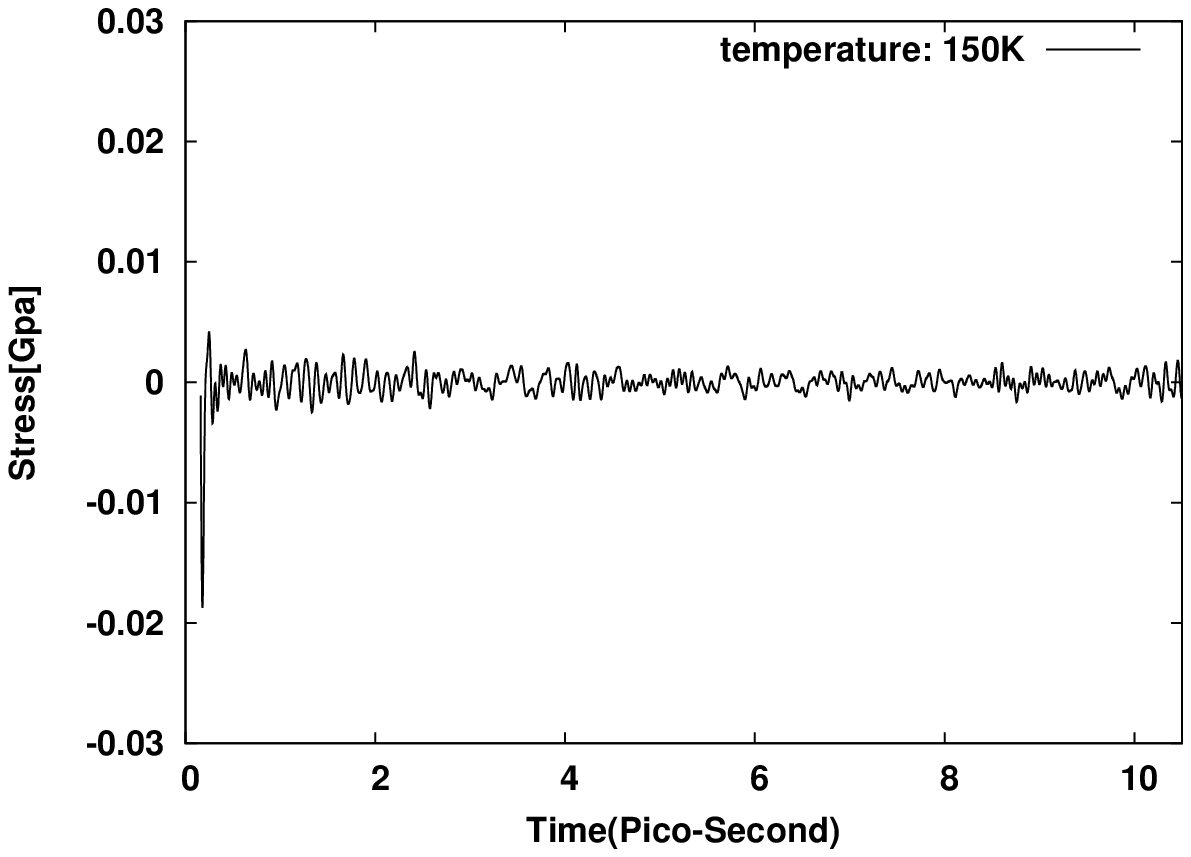}
\end{center}
\caption{Variation of $(\bs\sigma_{11}^V-{\det(\mb F)}^{-1}\mb F\cdot\mb P_{11}^V)$ and $(\widehat{\bs\sigma_{11}^V}-{\det(\mb F)}^{-1}\mb F\cdot\widehat{\mb P_{11}^V})$ at different temperatures. The first row is the original value, the second row is applied with temporal average with radius of $0.159$ pico-seconds. From left to right, the temperatures are 50K, 100K, 150K. }
\label{fig:v_k_sum}
\end{figure}

The first row of Figure \ref{fig:v_k_sum} shows the variation of $(\bs\sigma^V-{\det(\mb F)}^{-1}\mb F\cdot\mb P^V)$ at different temperatures. From the first row, we can see the value keeps varying around zero value despite the increasing temperature and pressure. However, the amplitude of the oscillations increases as the temperature increases.

The second row of Figure \ref{fig:v_k_sum} shows the variation of $(\bs\sigma^V-{\det(\mb F)}^{-1}\mb F\cdot\mb P^V)$ under a temporal average with radius of $0.159$ pico-seconds. We can see the amplitudes of oscillations decrease significantly after temporal averages are applied. However, we still see slightly larger amplitude of oscillations at higher temperature.

\begin{figure}[htbp]
\begin{center}
\includegraphics[scale=0.5]{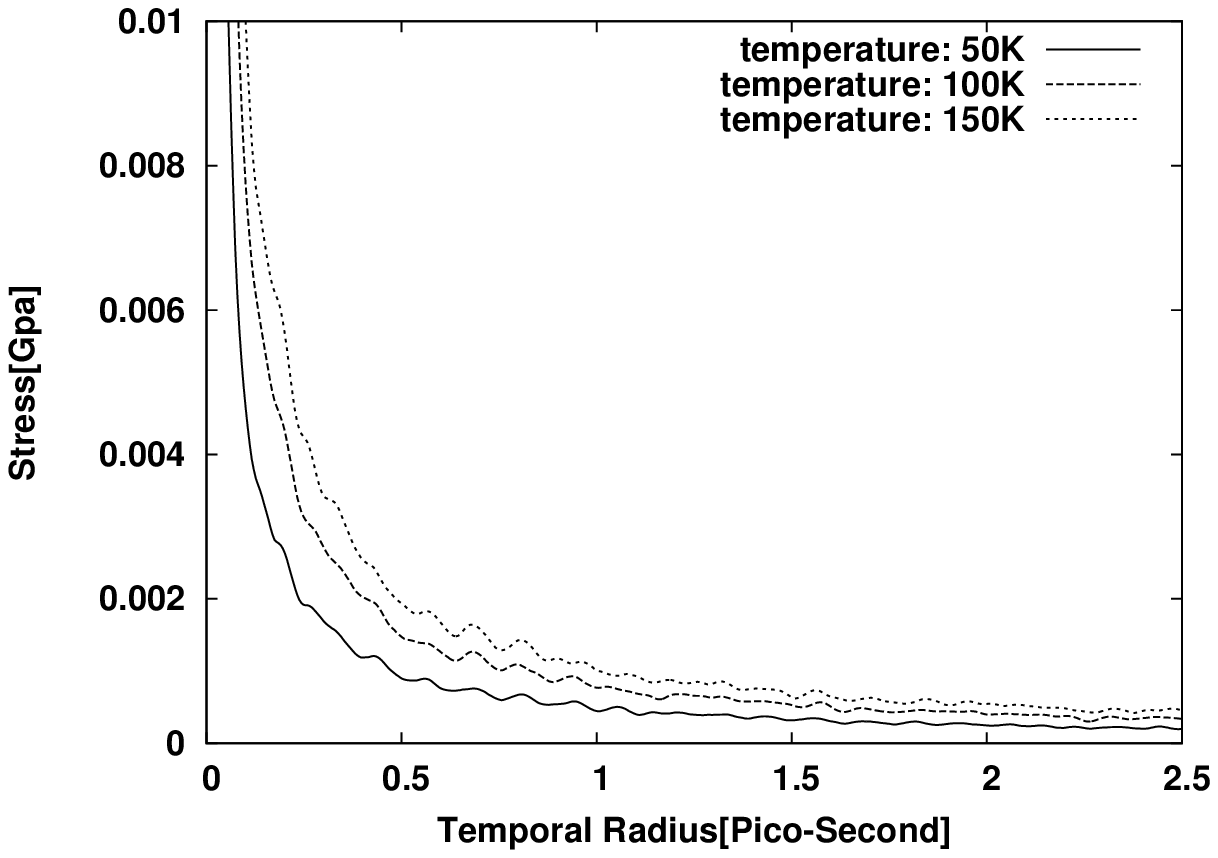}
\includegraphics[scale=0.5]{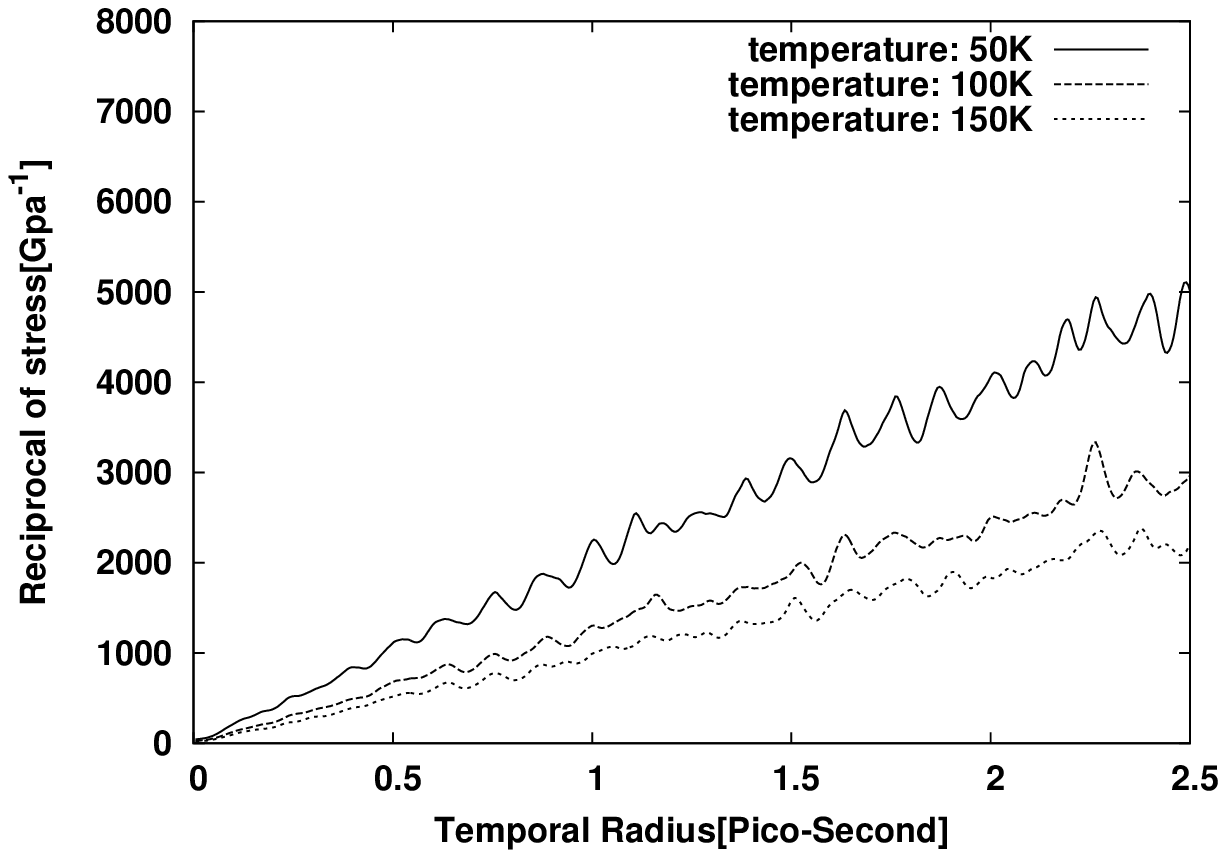}
\end{center}
\caption{The original (left) and reciprocal (right) value of
$|\widehat{\bs\sigma_{11}^V}-{\det(\mb F)}^{-1}\mb F\cdot\widehat{\mb P_{11}^V}|$ under different temporal average radius and temperatures.}
\label{fig:v_k_trtemp}
\end{figure}

To see more clear relationship between $(\bs\sigma^V-{\det(\mb F)}^{-1}\mb F\cdot\mb P^V)$ and the system temperature and temporal kernel radius. We draw Figure \ref{fig:v_k_trtemp}. The values are first taken at each time-step, then we take the mean absolute value of them. This process is repeated under different temporal radius and temperatures. We also draw the reciprocal of the value, thus the dependence on temporal radius can be more clearly observed.

The left figure in Figure \ref{fig:v_k_trtemp} shows the original values. We can see when temporal radius increases, all the three curves decrease, and the curve representing for the higher temperature has higher values over whole. In the right figure of Figure \ref{fig:v_k_trtemp}, the value in y-axis has been taken reciprocal value. We can see the three curves are nearly in linear growth. This shows that, the original data in the left figure are nearly in order $O(r_t^{-1})$. This order of decreasing is consistent with our analysis prediction.

The numerical experiments are consistent with our analysis in B of section \ref{sec:pk-cauchy-in-gIK}. Namely, the temporal average and temperature are two deciding factors that determine the difference between Cauchy and revised PK stress. The larger temporal radius will decrease their difference, while the higher temperature will increase the difference.

\section{Conclusion}\label{sec:conclu}

\begin{enumerate}
\item
The g-IK stress has intrinsic difference with the time-averaged stress.
This difference can be presented in two aspects. The first is when the space-time kernel is inseparable. In this case, the g-IK stress gives different domain of spacial average at different time, while the time-averaged Hardy stress has the same domain of spacial average at all instant of time. The second is when the kernel is space-time separable. In this case,
the difference can not be neglected when the spacial radius is small and the temporal radius is large. However, a large enough spacial radius or a small enough temporal radius can guarantee their difference vanishing.
\item
The connections between several versions of stress can be rebuilt in the framework of g-IK formulism. The first is when the spacial radius is large enough, the virial stress can be regarded as g-IK stress when the space-time kernel has uniform weighting form. The second is on
the classical relationship between Cauchy and first Piola-Kirchhoff stress (\ref{eq:cla_pk_cc}). We show the relationship holds only when the system temperature is finite and the temporal radius is large enough, and the dependence of the difference between PK and revised Cauchy stress on the temporal average radius is about $O(r_t^{-1})$.
\end{enumerate}

\bibliographystyle{plain}

\end{document}